\documentclass[aps,prd,superscriptaddress,showpacs,preprintnumbers,10pt,floatfix]
{revtex4}
\usepackage{graphicx,color,amsmath}

\newcommand{\be}{\begin{equation}}
\newcommand{\ee}{\end{equation}}
\newcommand{\bea}{\begin{eqnarray}}
\newcommand{\eea}{\end{eqnarray}}

\newcommand{\bfk}{\mbox{\boldmath $k$}}

\def\kt{k_\perp}
\newcommand{\bfp}{\mbox{\boldmath $p$}}
\def\bpo{{\bfp}_{\perp 1}}
\def\bpt{{\bfp}_{\perp 2}}

\newcommand{\bfq}{\mbox{\boldmath $q$}}
\newcommand{\bfP}{\mbox{\boldmath $P$}}

\def\ppo{p_{\perp 1}}
\def\ppt{p_{\perp 2}}
\def\pp{p_\perp}

\newcommand{\ua}{\uparrow}

\def\avk{\langle k_\perp ^2\rangle}
\def\avp{\langle p_\perp ^2\rangle}
\def\avk{\langle k_\perp ^2\rangle}
\def\avp{\langle p_\perp ^2\rangle}
\def\avPT{\langle P_T^2\rangle}

\def\T{_{_T}}
\def\C{_{_C}}

\newcommand{\deltatilde}{\tilde{\Delta} ^N D}

\def\lsim{\mathrel{\rlap{\lower4pt\hbox{\hskip1pt$\sim$}}\raise1pt\hbox{$<$}}}
\def\gsim{\mathrel{\rlap{\lower4pt\hbox{\hskip1pt$\sim$}}\raise1pt\hbox{$>$}}}
\def\nostrocostruttino#1\over#2{\mathrel{\mathop{\kern 0pt \rlap
{\hbox{$#1$}}} \hbox{\kern-.135em $#2$}}}

\newcommand{\BELLEZ}{\mbox{Belle-$z_1 z_2$}}
\newcommand{\BABARZ}{\mbox{BaBar-$z_1 z_2$}}
\newcommand{\BABARPTZ}{\mbox{BaBar-$P_{1T}$}}

\begin{document}
\title{Collins functions for pions from SIDIS and new $e^+e^-$ data: \\
a first glance at their transverse momentum dependence}

\author{M.~Anselmino}
\affiliation{Dipartimento di Fisica, Universit\`a di Torino,
             Via P.~Giuria 1, I-10125 Torino, Italy}
\affiliation{INFN, Sezione di Torino, Via P.~Giuria 1, I-10125 Torino, Italy}
\author{M.~Boglione}
\affiliation{Dipartimento di Fisica, Universit\`a di Torino,
             Via P.~Giuria 1, I-10125 Torino, Italy}
\affiliation{INFN, Sezione di Torino, Via P.~Giuria 1, I-10125 Torino, Italy}
\author{U.~D'Alesio}
\affiliation{Dipartimento di Fisica, Universit\`a di Cagliari,
             I-09042 Monserrato (CA), Italy}
\affiliation{INFN, Sezione di Cagliari,
             C.P.~170, I-09042 Monserrato (CA), Italy}
\author{J.O.~Gonzalez~Hernandez}
\affiliation{Dipartimento di Fisica, Universit\`a di Torino,
             Via P.~Giuria 1, I-10125 Torino, Italy}
\affiliation{INFN, Sezione di Torino, Via P.~Giuria 1, I-10125 Torino, Italy}
\author{S.~Melis}
\affiliation{Dipartimento di Fisica, Universit\`a di Torino,
             Via P.~Giuria 1, I-10125 Torino, Italy}
\author{F.~Murgia}
\affiliation{INFN, Sezione di Cagliari,
             C.P.~170, I-09042 Monserrato (CA), Italy}
\author{A.~Prokudin}
\affiliation{Division of Science,
                   Penn State Berks,
                   Reading, PA 19610, USA}

%
\date{\today}

\begin{abstract}
New data from Belle and BaBar Collaborations on azimuthal asymmetries,
measured in $e^+e^-$ annihilations into pion pairs at $Q^2=112$
GeV$^2$, allow to take the first, direct glance at the $p_\perp$ dependence
of the Collins functions, in addition to their $z$ dependence.
These data, together with available Semi-Inclusive Deep Inelastic Scattering
(SIDIS) data on the Collins asymmetry,
are simultaneously analysed in the framework of the generalised parton model
assuming two alternative $Q^2$ evolution schemes and exploiting two different
parameterisations for the Collins functions. The corresponding results
for the transversity distributions are presented.
Analogous data, newly released by the BESIII Collaboration, on $e^+e^-$
annihilations into pion pairs at the lower $Q^2$ of $13$ GeV$^2$,
offer the possibility to explore the sensitivity of these azimuthal correlations
on transverse momentum dependent evolution effects.
\end{abstract}



\maketitle

\section{\label{Intro} Introduction}

Our knowledge of the 3-dimensional partonic structure of the nucleon in
momentum space is encoded, at leading-twist, in eight Transverse Momentum
Dependent Parton Distribution Functions (TMD-PDFs). They depend on two
variables, the light-cone momentum fraction, $x$, of the parent nucleon's momentum
carried by a parton and the parton transverse momentum, $\bfk_\perp$, with
respect to the direction of the nucleon's motion. At a low resolution scale $Q^2$
the transverse momentum  $\bfk_\perp$ may be associated with the intrinsic motion
of confined partons inside the nucleon. For polarised nucleons and partons
there is a further dependence on the spins of the nucleon and the parton.
In addition, the QCD radiation of gluons induces a dependence on the scale
$Q^2$ at which the nucleon is being explored.

Similarly, the hadronisation process of a parton into the final hadron is
encoded in the Transverse Momentum Dependent parton Fragmentation Functions
(TMD-FFs), which, in addition to spin   depend on
the light-cone momentum fraction, $z$, of the fragmenting parton carried by
the hadron and the hadron transverse momentum, $\bfp_\perp$,  with respect to
the parton direction. For final spinless or unpolarised
hadrons there are, at leading-twist, two independent TMD-FFs.

So far, among the polarised leading twist TMD-PDFs and TMD-FFs, the Sivers
distribution~\cite{Sivers:1989cc,Sivers:1990fh} and the Collins fragmentation
function~\cite{Collins:1992kk} have clearly shown their non negligible effects
in several different experimental measurements. The former describes the
correlation between the intrinsic momentum $\bfk_\perp$ of unpolarised partons
and the parent nucleon transverse spin; as such, it must be related to parton
orbital angular momentum. The latter describes the correlation between
the transverse spin of a fragmenting quark and the transverse momentum
$\bfp_\perp$ of the final produced hadron, typically a pion, with respect
to the quark direction; as such, it reveals fundamental properties of the
hadronization process. This paper is devoted to the study of the Collins
functions.

The Collins fragmentation function can be studied in Semi-Inclusive Deep
Inelastic Scattering (SIDIS) experiments,
where it appears convoluted with the transversity distribution, and where,
being dependent on $\bfp_\perp$, it induces a typical azimuthal modulation,
the so-called Collins asymmetry. Clear signals of this
asymmetry were observed experimentally, see Refs.~\cite{Airapetian:2010ds,
Adolph:2012sn,Adolph:2014zba}.
The Collins fragmentation functions also induce azimuthal angular correlations
between hadrons produced in opposite jets in $e^+e^-$
annihilation~\cite{Boer:1997mf, Abe:2005zx}. Consequently, a simultaneous
analysis of SIDIS and $e^+e^-$ data allows the combined extraction of the
transversity distribution and the Collins FFs~\cite{Anselmino:2007fs,
Anselmino:2008jk, Anselmino:2013vqa}.

Very recently, new data on the $e^+e^- \to h_1\,h_2\,X$ process have been
published by the BaBar Collaboration, focusing on their $z$ and $p_\perp$
dependence~\cite{TheBABAR:2013yha}. It is the first direct
measurement of the transverse momentum dependence of an asymmetry, in
$e^+e^-$ processes, related to TMD functions. BaBar data benefit from
very high statistics and offer, in addition to the $z_1,z_2$ distributions,
data on the $A_{12}$ asymmetry in bins of $(z_1,z_2,\ppo,\ppt)$
and in bins of $\ppo$ and $\ppt$, where $\ppo$ and $\ppt$ are the transverse
momenta of the final hadrons with respect to the thrust axis.
Moreover, BaBar measures the $A_{0}$ asymmetry as a function of $P_{1T}$,
the transverse momentum of the final hadron $h_1$ with respect to the plane
which contains both the $e^+e^-$ pair and the other final hadron, $h_2$, in the
$e^+e^-$ c.m. frame. Information on the transverse momentum dependence of
the asymmetries thus allows a first glance at the dependence of the Collins
FFs on the transverse momentum, $p_\perp$.

The explicit dependence of the TMDs  on their corresponding momentum fractions
$x$ or $z$ is relatively easy to access, as most measured observables
(cross sections, multiplicities, asymmetries) are given as functions of $x$ or
$z$, although still in a limited range.
Instead, the transverse momentum dependence is much more involved, as $k_\perp$ 
and $p_\perp$ are never observed  directly
but only through convolutions. Asymmetries alone are not sufficient for a
complete study of the Collins transverse momentum dependence,
as they require the knowledge of the unpolarised TMD fragmentation functions,
which appear in the denominator of the asymmetry.
Information on the unpolarised FFs have historically been extracted from
SIDIS processes where, unfortunately, the $k_\perp$ and $p_\perp$ dependences
are strongly correlated and cannot be disentangled unambiguously.
For a direct extraction of the $p_\perp$ dependence of the unpolarised
FFs one would need to measure, for example,
transverse momentum dependent cross sections or multiplicities in
$e^+e^-\to h_1 \, h_2 \, X$ processes, which would,
finally, allow the extraction of the $p_\perp$ dependence of
the Collins function from $e^+e^-$ asymmetries. Although the present study cannot
deliver an absolute determination of the Collins function, our analysis of
the new BaBar measurements allows to obtain the relative TMD behaviour of
the Collins function with respect to that of the unpolarised TMD--FF.


In this paper we adopt a phenomenological model for TMD--PDFs and FFs
in a scheme where the cross section is written as the convolution of two TMDs
with the corresponding partonic cross section. Moreover, we assume that the
TMD longitudinal and transverse degrees of freedom factorize.
The $z$-dependent part of our TMDs evolves in $Q^2$ while the transverse momentum
dependent part is $Q^2$ independent.
This model, sometimes called Generalized Parton Model (GPM), has proven to
work surprisingly well, allowing to describe a wide variety of observables:
from the SIDIS unpolarised multiplicities~\cite{Schweitzer:2010tt,
Anselmino:2013lza,Signori:2013mda}, to SIDIS Sivers and Collins
effects~\cite{Anselmino:2008sga,Anselmino:2012aa,Anselmino:2013vqa}
up to the most intriguing spin asymmetries in inclusive hadron
production~\cite{Anselmino:2012rq,Anselmino:2013rya}.

Proper treatment of TMDs would require the use of TMD
evolution~\cite{Collins:2011qcdbook}. In fact, one expects that, as $Q^2$ grows, 
gluon radiations will change the functional form of the $k_\perp$ and
$p_\perp$ dependence: in particular, the widths of the TMDs will generically
grow with $Q^2$. The corresponding evolution equations are the so-called
Collins-Soper (CS) equations~\cite{Collins:1981uk,Collins:1981uw}. Recently,
evolution equations have been formulated for unpolarised TMD functions
directly~\cite{Collins:2011qcdbook,Aybat:2011zv,Echevarria:2012pw,
Echevarria:2014rua}. 
Polarised TMDs, in particular the Collins FFs, were shown to have similar
evolution equations~\cite{Echevarria:2014rua} and the first analysis
of the SIDIS and $e^+e^-$ data including TMD evolution was presented in
Ref.~\cite{Kang:2015msa}. The results of Ref.~\cite{Kang:2015msa} are similar
to the GPM model results published in Ref.~\cite{Anselmino:2013vqa}.

The TMD approach is valid in the region in which $q_T \ll Q$, where $q_T \simeq P_T/z$
and $Q^2$ are the transverse momentum and the virtuality of the probing photon,
respectively. Available SIDIS data cover the region from low to
moderate $Q^2$. For instance the average values of $Q^2$ of the SIDIS data
considered in the present analysis are between $2.4$ and $3.2$ GeV$^2$, while
the typical transverse momentum $P_T$ of the final hadron is between $0.1$ and
$1.5$ GeV. Clearly, in this region, it is difficult to guarantee $q_T \ll Q$.
It is then crucial to test the validity of the TMD approach in this range of
$Q^2$ and $q_T$ by comparing our results and those obtained by applying a
TMD evolution scheme~\cite{Kang:2015msa} to the available experimental data.

In principle $e^+e^-$ Collins asymmetry data, which correspond to a much
larger $Q^2$, allow the application of the TMD evolution scheme in its range
of validity. However, the observables we are analyzing are, in general,
ratios or double ratios of cross sections, where strong cancellations of TMD evolution effects can occur.
Therefore, we have to understand whether soft gluon emissions, typical of TMD
evolution, affect the Collins asymmetries, and whether we can unambiguously 
observe any explicit $Q^2$ dependence in the presently available data. This 
might also help to better determine the universal, non-perturbative part 
of TMD evolution~\cite{Aidala:2014hva,Collins:2014jpa}.
Having no $Q^2$ evolution in the transverse momentum distribution,
our model could be considered as a benchmark for these kind of studies.

Almost at completion of our paper new results from the BESIII Collaboration
have appeared~\cite{Ablikim:2015sma}. They definitely confirm the need of
having non vanishing Collins functions; in addition, they present the very
interesting feature of being at much lower $Q^2$ values with respect to
Belle and BaBar data. We do not include them in our fitting
procedure, but rather we will compare our determination of the Collins 
functions with these new results, and explore the sensitivity
of these azimuthal correlations on $Q^2$ dependent effects.

The purpose of this paper is twofold. First, we would like to test the GPM
against the new $e^+e^-$ data, both from BaBar and BESIII Collaborations, and
see whether the newest data put limitations on the region of applicability
of our model. Second, we would like to study the $p_\perp$ dependence of the
pion Collins functions.

We only consider here pion production. The BaBar Collaboration has
recently also measured azimuthal correlations for pion-kaon and
kaon-kaon pairs produced in $e^+e^-$ annihilations~\cite{Aubert:2015hha}.
They allow the first ever extraction of the kaon Collins function and will
be considered in a forthcoming paper.

The paper is organized as follows. In Section~\ref{Form} we briefly recall
the formalism used in our analysis, while in Section~\ref{Mod} we discuss
our best fit of Belle~\cite{Seidl:2008xc,Seidl:2012er}, BaBar~\cite{TheBABAR:2013yha},
HERMES~\cite{Airapetian:2010ds} and COMPASS~\cite{Martin:2013eja} results
and present our extraction of the valence quark transversity distributions 
and of the pion Collins functions.
In Section~\ref{Collins-evo} we study how our choice of parameterisation
for the Collins function affects the results of our fit and study its
dependence on the chosen evolution scheme.
The newly released, low energy, BESIII data will be discussed in
Section~\ref{BESIII}. Final comments, including some considerations on
the role of TMD evolution in phenomenological analyses of asymmetries,
will be given in Section~\ref{Com}, together with our conclusions.

\section{\label{Form} Formalism}

Our strategy for the extraction of the TMD transversity and Collins functions
is based on a simultaneous best fit of SIDIS and $e^+e^-\to h_1 \, h_2\, X$
experimental data. We only summarise here the basic formalism used throughout
the paper; all details can be found in Refs.~\cite{Anselmino:2007fs,
Anselmino:2011ch, Anselmino:2013vqa} to which we refer for notations,
kinematical variables, and for the definition of the azimuthal angles
which appear in the following equations.

\subsection{SIDIS}

In SIDIS processes, at ${\cal O}(k_\perp/Q)$, the $\sin(\phi_h+\phi_S)$ moment
of the measured spin asymmetry $A_{UT}$~\cite{Anselmino:2011ch, Bacchetta:2006tn},
is proportional to the spin dependent part of the fragmentation
function of a transversely polarised quark, encoded in the Collins function,
$\Delta^N\!D_{h/q^\uparrow}(z,p_\perp) = (2 \, p_\perp/z \, m_h) \,
H_1^{\perp q}(z,\pp)$~\cite{Bacchetta:2004jz}, convoluted with the TMD transversity
distribution $\Delta_T q(x,k_\perp)$~\cite{Anselmino:2007fs}:
\be
A^{\sin (\phi_h+\phi_S)}_{UT} = \label{sin-asym}
\frac{\displaystyle  \sum_q e_q^2  \! \! \int \! \!{d\phi_h \, d\phi_S \, d^2
\bfk _\perp}\,\Delta _T q (x,\kt) \,
\frac{d (\Delta {\hat \sigma})}{dy}\,
\Delta^N D_{h/q^\ua}(z,\pp) \sin(\phi_S + \varphi +\phi_q^h)
\sin(\phi_h + \phi_S) } {\displaystyle \sum_q e_q^2 \, \int {d\phi_h
\,d\phi_S \, d^2 \bfk _\perp}\; f_{q/p}(x,k _\perp) \;
\frac{d\hat\sigma}{dy}\; \; D_{h/q}(z,p_\perp) } \> \cdot
\ee

The above equation further simplifies when adopting a Gaussian and
factorised parameterisation for the TMDs. In particular for the unpolarised
parton distribution and fragmentation functions we assume:
\bea
 f_{q/p}(x,k_\perp) & = &
 f_{q/p}(x)\;\frac{e^{-{k_\perp^2}/\avk}}{\pi\avk} \label{funp} \,,\\
 D_{h/q}(z,\pp)& =& D_{h/q}(z)\;\frac{e^{-\pp^2/\avp}}{\pi\avp}
 \label{dunp}\;\cdot
\eea

For the integrated parton distribution and fragmentation functions,
$f_{q/p}(x)$ and $D_{h/q}(z)$, we use respectively the GRV98LO PDF
set~\cite{Gluck:1998xa} and the De Florian, Sassot and Stratmann (DSS) FF set~\cite{deFlorian:2007hc}.
This choice is dictated by the fact that GRV98LO is the only PDF set
with an initial scale, $Q_0$, low enough to accommodate all HERMES
data points, including those at the lowest values of $Q^2$.
We have checked that different choices of distribution and fragmentation
function sets hardly influence the outcome of our analysis.
The Gaussian widths are fixed to the values obtained by fitting HERMES SIDIS
multidimensional multiplicities in Ref.~\cite{Anselmino:2013lza}:
\be
\avk =0.57 \> {\rm GeV}^2 \quad\quad\quad \avp = 0.12 \> {\rm GeV}^2 \>;
\label{widths}
\ee
notice that these values were obtained using the unpolarised CTEQ6LO
PDFs~\cite{Pumplin:2002vw}, rather than the GRV98LO PDFs, adopted here;
again, we have explicitly checked that using the GRV98LO PDFs in fitting
the multidimensional multiplicities would not change the above results.

These values are different from those obtained and adopted in  previous
analyses ~\cite{Anselmino:2007fs,
Anselmino:2011ch, Anselmino:2013vqa}. The determination
of the separate values of $\avk$ and $\avp$ from SIDIS data is still rather
uncertain, and we have preferred here to choose the most recently obtained values,
which give a good fit~\cite{Anselmino:2013lza} of the unpolarised multiplicities.


{F}or the transversity distribution, $\Delta_T q(x, k_\perp)$, and
the Collins FF, $\Delta^N\! D_{h/q^\uparrow}(z,\pp)$, we adopt the
following factorised shapes~\cite{Anselmino:2007fs}:
\bea
\Delta_T q(x, k_\perp; Q^2) &=& \Delta_T q(x, Q^2)
\> \frac{e^{-{k_\perp^2}/{\avk\T}}}{\pi \avk \T}\,, \label{tr-funct}\\
\Delta^N \! D_{h/q^\uparrow}(z,\pp; Q^2) &=&  \tilde{\Delta} ^N D_{h/q^\ua}(z, Q^2)
\> h(\pp)\,\frac{e^{-\pp^2/{\avp}}}{\pi \avp}\,,
\label{coll-funct}
\eea
where $\Delta_T q(x)$ is the integrated transversity distribution and
$\tilde{\Delta} ^N D_{h/q^\ua}(z)$ is the $z$-dependent part of the Collins
function. In order to easily implement the proper positivity bounds, these
functions are written, at the initial scale $Q_0^2$, as~\cite{Anselmino:2007fs}
\bea
&&\Delta_T q(x,Q_0^2) =  {\cal N}^{\T}_q(x,Q_0^2)\,
\frac{1}{2}\,[f_{q/p}(x,Q_0^2)+\Delta q(x,Q_0^2)]
\label{coll-transv} \\
&&\tilde{\Delta} ^N D_{h/q^\ua}(z,Q_0^2) = 2 \,
{\cal N}^{\C}_{q}(z,Q_0^2)\,D_{h/q}(z,Q_0^2) \>.
\label{coll-D}
\eea
They are then evolved up to the proper value of $Q^2$. For
$\Delta_T q(x, Q^2)$ we employ a transversity DGLAP kernel and the evolution
is performed by an appropriately modified Hoppet code~\cite{Salam:2008qg}.
The Soffer bound is built in by using the GRV98LO~\cite{Gluck:1998xa} and
GRSV2000~\cite{Gluck:2000dy} PDF sets at the input scale of $Q_0^2=1$ GeV$^2$,
with $\alpha_s(Q_0)=0.44$ calculated according to the GRV98 LO scheme. In this
analysis, we use a simplified model which implies no $Q^2$ dependence in the $\pp$ distribution.
As the Collins function in our parameterisation is proportional to
the unpolarised fragmentation function, see Eqs.~\eqref{coll-funct} and
\eqref{coll-D}, we assume that the only scale dependence is contained in
$D(z,Q^2)$, which is evolved with an unpolarised DGLAP kernel, while
${\cal N}^{\C}_{q}$ does not evolve with $Q^2$. This is equivalent to assuming
that the ratio $\tilde{\Delta}^N D(z,Q^2)/D(z,Q^2)$ is constant in $Q^2$.
Throughout the paper, we will refer to this choice as the ``standard''
evolution scheme.

The function $h(\pp)$, defined as~\cite{Anselmino:2007fs}
\be
h(\pp)=\sqrt{2e}\,\frac{p_\perp}{M_{C}}\,e^{-{p_\perp^2}/{M_{C}^2}}
\label{hpcollins}\,,
\ee
allows for a possible modification of the $\pp$ Gaussian width of the
Collins function with respect to the unpolarised FF; for the TMD
transversity distribution, instead, we assume the same Gaussian width
as for the unpolarised TMD, $\avk\T = \avk$.

We parameterise ${\cal N}^{\T}_q(x)$ as
\be
{\cal N}^{\T}_q(x)=N^{\T}_q \,x^{\alpha}(1-x)^\beta\,\,
\frac{(\alpha+\beta)^{\alpha+\beta}}{\alpha^\alpha \beta^\beta}
\qquad (q = u_v,d_v)
\ee
where $-1\le N^{\T}_q\le +1$, $\alpha$ and $\beta$ are free parameters of
the fit. Thus, the transversity distributions depend on a total of 4 parameters
($N^{\T}_{u_v}, N^{\T}_{d_v}, \alpha, \beta$).

{For} the Collins function, as in previous papers~\cite{Anselmino:2007fs,
Anselmino:2013vqa}, we distinguish between favoured and disfavoured
fragmentations. The favoured contribution is parameterised as
\be
\mathcal{N}^{\C}_{\rm fav}(z)=N^{\C}_{\rm fav} \,z^{\gamma}(1-z)^\delta\,\,
\frac{(\gamma+\delta)^{\gamma+\delta}}{\gamma^\gamma \delta^\delta},
\label{std-fav}
\ee
where $-1\le N^{\C}_{\rm fav}\le +1$, $\gamma$ and $\delta$ are free parameters
of the fit. Differently from what we did in the past, we do not assume
the same functional shape for favoured and disfavoured Collins functions.
In a first attempt we chose for $\mathcal{N}^{\C}_{\rm dis}$ a parameterisation
analogous to that shown in Eq.~\eqref{std-fav}, letting the fit free to
choose different $\gamma$ and $\delta$ parameters. It turned out that, for the
disfavoured Collins function, the best fit values of $\gamma$ and $\delta$
were very close or compatible with zero. One should also notice that, 
with the presently available SIDIS and $e^+e^-$ data, the disfavoured Collins function is
largely undetermined. Consequently, also in order to reduce the number of
parameters, we simply choose
\be
\mathcal{N}_{\rm dis}^{\C}(z) = N^{\C}_{\rm dis}\,.
\label{std-dis}
\ee
Thus, we have a total of five free parameters for the Collins functions
($M_C, N^{\C}_{\rm fav}, N^{\C}_{\rm dis}, \gamma, \delta)$.
Notice that, although present data are still unable to tightly constrain
the disfavoured Collins function, it clearly turns out that choosing
independent parameterisations for $\mathcal{N}_{\rm fav}^{\C}(z)$ and
$\mathcal{N}_{\rm dis}^{\C}(z)$ definitely improves the quality of the fit.

Using Eqs.~\eqref{funp},~\eqref{dunp},~\eqref{tr-funct},~\eqref{coll-funct}
into Eq.~\eqref{sin-asym} we obtain the following expression for
$A^{\sin (\phi_h+\phi_S)}_{UT}$:
\be
A^{\sin (\phi_h+\phi_S)}_{UT} =
\frac{\displaystyle \sqrt{2e} \, \frac{P_T}{M_C}\,
 \frac{\avp ^2 \C}{\avp}
\, \frac{e^{-P_T^2/\avPT \C}}{\avPT ^2 \C}
\frac{1-y}{s x y^2}\, \sum_q e_q^2 \;
\Delta_T q(x)\;
\tilde{\Delta} ^N D_{h/q^\ua}(z)}
{ \displaystyle \frac{e^{-P_T^2/\avPT}}{\avPT} \,
\frac{[1+(1-y)^2]}{s x y^2}\,
 \sum_q e_q^2 \, f_{q/p}(x)\; D_{h/q}(z)}\;,
\label{sin-asym-final-1}
\ee
with
\bea
 \avp\! \C= \frac{M_{C}^2 \, \avp}{M_{C}^2 +\avp} &&\hspace*{1.5cm}
 \avPT_{_{\!(C)}} =\avp_{_{\!(C)}} +z^2\avk \,.
\eea



\subsection{$e^+e^-\to h_1 h_2\, X$ processes}

Independent information on the Collins functions can be
obtained in unpolarised $e^+e^-$ processes, by looking at the azimuthal
correlations of hadrons produced in opposite jets~\cite{Boer:1997mf}.
The Belle Collaboration~\cite{Abe:2005zx,Seidl:2008xc,Seidl:2012er} and,
more recently, the BaBar Collaboration~\cite{TheBABAR:2013yha}
have measured azimuthal hadron-hadron correlations for inclusive charged
pion production in $e^+e^-\to \pi \, \pi \, X$ processes, which, involving
the convolution of two Collins functions, can be interpreted as a direct
measure of the Collins effect.

Two methods have been adopted in the experimental analysis of the Belle and
BaBar data~\cite{Boer:1997mf,Anselmino:2007fs,Seidl:2008xc,TheBABAR:2013yha}:
\begin{enumerate}
\item
In the ``thrust-axis method'' the jet thrust axis, in the $e^+e^-$ c.m. frame,
fixes the $\hat z$ direction and the $e^+e^-\to q \, \bar q$ scattering defines
the $\widehat{xz}$ plane; $\varphi_1$ and $\varphi_2$ are the azimuthal angles
of the two hadrons around the thrust axis, while $\theta$ is the angle between
the lepton direction and the thrust axis. In this reference frame, with
unpolarised leptons, the cross section can be written as~\cite{Anselmino:2007fs}:
\begin{widetext}
\bea
\frac{d\sigma ^{e^+e^-\to h_1 h_2 X}}{dz_1\,dz_2\,\ppo \, d\ppo \, \ppt \,
d\ppt\,d\cos\theta \, d(\varphi_1 + \varphi_2)}
&=&
 \frac{3\pi^2\alpha^2}{s} \, \sum _q e_q^2 \, \Big\{
 (1+\cos^2\theta)\,D_{h_1/q}(z_1,\ppo)\,D_{h_2/\bar q}(z_2,\ppt)
 \Big. \nonumber \\ &+&
 \Big. \frac{1}{4}\,\sin^2\theta\,\Delta ^N D _{h_1/q^\ua}(z_1,\ppo)\,
 \Delta ^N D _{h_2/\bar q^\ua}(z_2,\ppt)\,\cos(\varphi_1 + \varphi_2)\Big\}.  \nonumber \\~
 \label{belle}
\eea
\end{widetext}

Until very recently, only data on the $z$ dependence were available,
while $\ppo$ and $\ppt$ were integrated out. However, in 2014 the BaBar
Collaboration has released a new analysis in which multidimensional
data are presented~\cite{TheBABAR:2013yha}. This represents the first
direct measurement of the dependence of the Collins function on the
intrinsic transverse momenta $\ppo$ and $\ppt$.

By normalizing Eq.~(\ref{belle}) to the azimuthal averaged cross section,
\bea
\langle d\sigma \rangle &\equiv& \frac{1}{2\pi} \,
\frac{d\sigma^{e^+e^-\to h_1 h_2 X}}{dz_1\,dz_2 \,
\ppo \, d\ppo \, \ppt \, d\ppt \, d\cos\theta} \nonumber\\
&=&
\frac{3 \pi^2 \alpha^2}{s} \, \sum _q e_q^2 \,
 (1+\cos^2\theta)\,D_{h_1/q}(z_1,\ppo)\,D_{h_2/\bar q}(z_2,\ppt) \,, \label{phiav}
\eea
one has
\bea
R_{12}(z_1,z_2,\ppo,\ppt,\theta,\varphi_1 + \varphi_2) &\equiv& \frac{1}
{\langle d\sigma \rangle} \>
\frac{d\sigma ^{e^+e^-\to h_1 h_2 X}}{dz_1\,dz_2\,\ppo \, d\ppo\,\ppt \,
d\ppt\,d\cos\theta \, d(\varphi_1 + \varphi_2)} \nonumber\\
&=& 1 + \frac{1}{4}\,\frac{\sin^2\theta}{1+\cos^2\theta}\,
\cos(\varphi_1+\varphi_2)\, \nonumber \\
&\times&
\frac{\sum_q e^2_q \, \Delta ^N D_{h_1/q^\ua}(z_1,\ppo)\,
\Delta ^N D_{h_2/\bar q^\ua}(z_2,\ppt)}{\sum_q e^2_q D _{h_1/q}(z_1,\ppo)\,
D _{h_2/\bar q}(z_2,\ppt)}\,\cdot \label{A12g}
\eea

To eliminate false asymmetries, the Belle and BaBar Collaborations consider the
ratio of unlike-sign ($\pi^+\pi^-$ + $\pi^-\pi^+$) to like-sign
($\pi^+\pi^+$ + $\pi^-\pi^-$) or charged
($\pi^+\pi^+$ + $\pi^+\pi^- + \pi^-\pi^+$ + $\pi^-\pi^-$) pion pair production,
denoted respectively with indices $U$, $L$ and $C$.
For example, in the case of unlike- to like-pair production, one has
\bea
\frac{R_{12}^U}{R_{12}^L} &=& \frac{1+\frac{1}{4}\,\cos(\varphi_1+\varphi_2)\,
  \frac{ \sin^2\theta }{ 1+\cos^2\theta }\,P_U}
       {1+\frac{1}{4}\,\cos(\varphi_1+\varphi_2)
  \frac{\sin^2\theta}{ 1+\cos^2\theta }\,P_L}
\nonumber \\
& \simeq &  1+\frac{1}{4}\,\cos(\varphi_1+\varphi_2)
  \frac{\sin^2\theta }{ 1+\cos^2\theta } \,
  (P_U-P_L )\nonumber \\
&\equiv& 1+\cos(\varphi_1+\varphi_2)\,A_{12}^{UL}(z_1,z_2,\ppo,\ppt,\theta)\,,
\label{R}
\eea
with
\bea
P_U \equiv
      \frac{(P_U)_N}{(P_U)_D} = \frac{\sum_q e^2_q \;
    [\Delta ^N D _{\pi^+/q^\ua}(z_1,\ppo)\,
     \Delta ^N D _{\pi^-/\bar q^\ua}(z_2,\ppt) +
     \Delta ^N D _{\pi^-/q^\ua}(z_1,\ppo)\,
     \Delta ^N D _{\pi^+/\bar q^\ua}(z_2,\ppt)]}
{\sum_q e^2_q \;[D _{\pi^+/q}(z_1,\ppo)\,D _{\pi^-/\bar q}(z_2,\ppt) +
                 D _{\pi^-/q}(z_1,\ppo)\,D _{\pi^+/\bar q}(z_2,\ppt)]} \,,\nonumber
\\~\label{P_U}\\
P_L \equiv
      \frac{(P_L)_N}{(P_L)_D} = \frac{\sum_q e^2_q \;
    [\Delta ^N D _{\pi^+/q^\ua}(z_1,\ppo)\,
     \Delta ^N D _{\pi^+/\bar q^\ua}(z_2,\ppt) +
     \Delta ^N D _{\pi^-/q^\ua}(z_1,\ppo)\,
     \Delta ^N D _{\pi^-/\bar q^\ua}(z_2,\ppt)]}
{\sum_q e^2_q \;[D _{\pi^+/q}(z_1,\ppo)\,D _{\pi^+/\bar q}(z_2,\ppt) +
                 D _{\pi^-/q}(z_1,\ppo)\,D _{\pi^-/\bar q}(z_2,\ppt)]}\,,\nonumber \\ ~\label{P_L}
\eea
\be
A_{12}^{UL}(z_1,z_2,\ppo,\ppt,\theta)=\frac{1}{4}\frac{ \sin^2\theta }
{ 1+\cos^2\theta }\,(P_U-P_L )\,.\label{A12UL}
\ee
Similarly, for $A_{12}^{UC}(z_1,z_2,\ppo,\ppt,\theta)$ we have
\be
A_{12}^{UC}(z_1,z_2,\ppo,\ppt,\theta)=\frac{1}{4}\frac{\sin^2\theta}
{ 1+\cos^2\theta }\,(P_U-P_C )\,,
\label{A12UC}
\ee
where
\be
P_C = \frac{(P_U)_N + (P_L)_N}{(P_U)_D + (P_L)_D} \>\cdot
\ee

Notice that, in order to obtain the $\pp$ integrated asymmetries where only
the $z_1,z_2$ dependence is preserved, in Eqs.~\eqref{P_U} and~\eqref{P_L} 
we first integrate numerators and denominators separately 
over $\ppo$ and $\ppt$, and {\it then} we take ratios.

As said before, for fitting purposes it is convenient to introduce favoured and
disfavoured fragmentation functions, that is (see Eqs.~(\ref{coll-funct})
and (\ref{coll-D})):
\bea
&& \frac{\Delta^ND_{\pi^+/u^\ua,\bar d^\ua}(z, \pp)}{D_{\pi^+/u, \bar d}(z)}
=  \frac{\Delta^ND_{\pi^-/d^\ua,\bar u^\ua}(z, \pp)}{D_{\pi^-/d, \bar u}(z)}
=  2 \, {\cal N}^{\C}_{\rm fav}(z)\> h(\pp) \,
   \frac{e^{-\pp^2/{\avp}}}{\pi \avp} \label{fav} \\
&& \frac{\Delta^ND_{\pi^+/d^\ua,\bar u^\ua}(z, \pp)}{D_{\pi^+/d, \bar u}(z)}
=  \frac{\Delta^ND_{\pi^-/u^\ua,\bar d^\ua}(z, \pp)}{D_{\pi^-/u, \bar d}(z)}
=  \frac{\Delta^ND_{\pi^\pm/s^\ua,\bar s^\ua}(z, \pp)}{D_{\pi^\pm/s, \bar s}(z)}
=  2 \, {\cal N}^{\C}_{\rm dis}(z)\> h(\pp) \,
   \frac{e^{-\pp^2/{\avp}}}{\pi \avp} \> \cdot \label{unf}
\eea

\item
In the ``hadronic-plane method'',
one of the produced hadrons ($h_2$ in our case) identifies the
$\hat z$ direction and the $\widehat{xz}$ plane is determined by the
lepton and the $h_2$ directions;
the other relevant
plane is determined by $\hat z$ and the direction of the other
observed hadron, $h_1$, at an angle $\phi_1$ with respect to the
$\widehat{xz}$ plane. Here $\theta_2$ is the angle between $h_2$ and the $e^+e^-$ direction.

In this reference frame, the elementary process $e^+e^- \to q \, \bar q$
does not occur in the $\widehat{xz}$ plane, and thus the helicity scattering
amplitudes involve an azimuthal phase $\varphi_2$.
The analogue of Eq.~(\ref{belle}) now reads
\bea
\frac{d\sigma ^{e^+e^-\to h_1 h_2 X}}
{dz_1\,dz_2\,d^2\bpo\,d^2\bpt\,d\cos\theta_2}&=&
 \frac{3\pi\alpha^2}{2s} \, \sum _q e_q^2 \, \Big\{
 (1+\cos^2\theta_2)\,D_{h_1/q}(z_1,\ppo)\,D_{h_2/\bar q}(z_2,\ppt)
\frac{}{} \Big.  \!\!\! \label{belle2} \\ & &
+ \Big. \frac{1}{4}\,\sin^2\theta_2\,\Delta ^N D _{h_1/q^\ua}(z_1,\ppo)\,
 \Delta ^N D _{h_2/\bar q^\ua}(z_2,\ppt)\,
\cos(2\varphi_2 + \phi_{q}^{h{_1}})\Big\}\,,\nonumber
\eea
where $\phi_{q}^{h_1}$ is the azimuthal angle of the detected hadron $h_1$
around the direction of the parent fragmenting quark, $q$. In other words,
$\phi_{q}^{h_1}$ is the azimuthal angle of $\bpo$ in the helicity frame of
$q$. It can be expressed in terms of the integration variables we are using,
$\bpt$ and $\bfP_{1T}$, the transverse momentum of the $h_1$ hadron.
At lowest order in $\pp/(z\sqrt{s})$ we have
\bea
&&\cos\phi_{q}^{h_1} = \frac{P_{1T}}{\ppo} \,
\cos(\phi_1-\varphi_2) - \frac{z_1}{z_2} \, \frac{\ppt}{\ppo} \,,\\
&&\sin\phi_{q}^{h_1}=
\frac{P_{1T}}{\ppo} \, \sin(\phi_1-\varphi_2) \;.
\eea
The integration over $\bpt$ is performed explicitly, using the parameterisation
of the Collins function given in Eq.~\eqref{coll-funct}, while, as $\bpo =
\bfP_{1} - z_1\bfq _1$, we can replace $d^2\bpo$ with $d^2\bfP _{1T}$.
We obtain
\be
\frac{d\sigma ^{e^+e^-\to h_1 h_2 X}}
{dz_1\,dz_2\,d^2\bfP_{1T}\,d\cos\theta_2} =
\frac{3\pi\alpha^2}{2s} \Big\{D_{h_1\,h_2} + N_{h_1\,h_2} \cos2\phi_1 \Big\}\,,
\ee
where
\bea
D_{h_1\,h_2} &=& (1+\cos^2\theta_2)\,\sum _q e_q^2 D_{h_1/q}(z_1)\,
D_{h_2/\bar q}(z_2) \, \frac{ \exp{\Big[-\frac{P^2_{1T}}
{\langle\tilde{p}_\perp^2\rangle}\Big]}}
{\pi \langle\tilde{p}_\perp^2\rangle }\,,\\
N_{h_1\,h_2} &=& \frac{1}{4}\,\frac{z_1\,z_2}{z_1^2+z_2^2}\,\sin^2\theta _2\,
\sum_q e^2_q \, \deltatilde_{h_1/q^\ua}(z_1)\,
\deltatilde_{h_2/\bar q^\ua}(z_2) \,
\frac{2e \, P_{1T}^2}{\tilde{M}^2_C+\langle \tilde{p}_\perp^2 \rangle}\,
\frac{ \exp{\Big[-\frac{P^2_{1T}}{\tilde{M}^2_C} -\frac{P^2_{1T}}
{\langle\tilde{p}_\perp^2\rangle}\Big]}}
{\pi \langle\tilde{p}_\perp^2\rangle }\,,
\eea
and
\be
\tilde{M}^2_C=M_C^2 \,\frac{(z_1^2 + z_2^2)}{z_2^2}\,,\hspace*{2cm}
\langle \tilde{p}_\perp^2 \rangle = \avp\,\frac{(z_1^2 + z_2^2)}{z_2^2}\,\cdot \hspace*{2cm}
\label{tilde-funct}
\ee
The unlike, like and charged combinations are
\bea
D^U = D_{\pi^+\,\pi^-} + D_{\pi^-\,\pi^+}&\hspace*{2cm}& N^U =
N_{\pi^+\,\pi^-} + N_{\pi^-\,\pi^+}\\
D^L = D_{\pi^+\,\pi^+} + D_{\pi^-\,\pi^-}&\hspace*{2cm}& N^L =
N_{\pi^+\,\pi^+} + N_{\pi^-\,\pi^-}\\
D^C = D^U+ D^L \qquad\quad\,\, &\hspace*{2cm}& N^C = N^U+ N^L \,,
\eea
so that
\be
P^{U,L,C}_0 = \frac{N^{U,L,C}}{D^{U,L,C}}\,,
\ee
and finally
\be
R^{U,L,C}_0 = 1+P^{U,L,C}_0 \, \cos(2\phi_1)\,.
\ee
As in the previous case, we can build ratios of unlike/like and unlike/charged asymmetries:
\be
\frac{R^{U}_0}{R^{L(C)}_0}= \frac{1+P^{U}_0  \, \cos(2\phi_1)}
{1+P^{L(C)}_0  \, \cos(2\phi_1)}
\simeq 1+(P^{U}_0 - P^{L(C)}_0)  \, \cos(2\phi_1)
\equiv 1 + \cos(2\phi_1)\, A_{0}^{UL(C)} \label{A0ulc}\,,
\ee
which can then be directly compared to the experimental measurements.
\end{enumerate}

\section{\label{Mod} Best fit of SIDIS and $e^+e^-$ data: transversity
distributions, Collins functions and comparison with data}

We can now perform a best fit of the data on $A_{UT}^{\sin(\phi_h + \phi_S)}$
from HERMES and COMPASS and of the data on $A_{0}^{UL,C}$, from the Belle and
BaBar Collaborations. As anticipated above, we will not exploit the
$A_{12}^{UL,C}$ data in our fit, but only use them as a consistency check of our
results. In our fit -- we shall refer to it as the ``reference" fit -- these
asymmetries, given in Eqs.~(\ref{sin-asym-final-1}) and (\ref{A0ulc}), are
expressed in terms of the transversity and the Collins functions, parameterised
as in Eqs.~\eqref{funp}--\eqref{std-dis}, and evolved according to the
``standard" evolution scheme (see comments after Eq.~\eqref{coll-D}).

The transversity and the Collins functions depend on the free parameters
$\alpha, \beta, \gamma, \delta, N^{\T}_q, N^{\C}_q$ and $M_C$. Following
Ref.~\cite{Anselmino:2007fs} we assume the exponents $\alpha, \beta$ and
the mass scale $M_C$ to be flavour independent. Here we consider the
transversity distributions only for $u$ and $d$ valence quarks (with the
two free parameters $N^{\T}_{u_v}$ and $N^{\T}_{d_v}$). The favoured Collins
function is fixed by the flavour independent exponents $\gamma$ and $\delta$,
and by $N^{\C}_{\rm fav}$, while the disfavoured Collins function is
determined by the sole parameter $N^{\C}_{\rm dis}$
(see comments before Eq.~(\ref{std-dis})). This makes a total
of 9 parameters, to be fixed with a best fit procedure.
Notice that while in the present analysis we can safely neglect
any flavour dependence of the parameter $\beta$ (which is anyway very loosely
constrained by SIDIS data), this issue could play a significant role in other
studies, like the determination of the tensor charge~\cite{Anselmino:2012rq}.

\begin{figure}[h!t]
\includegraphics[width=9.9truecm,angle=0]{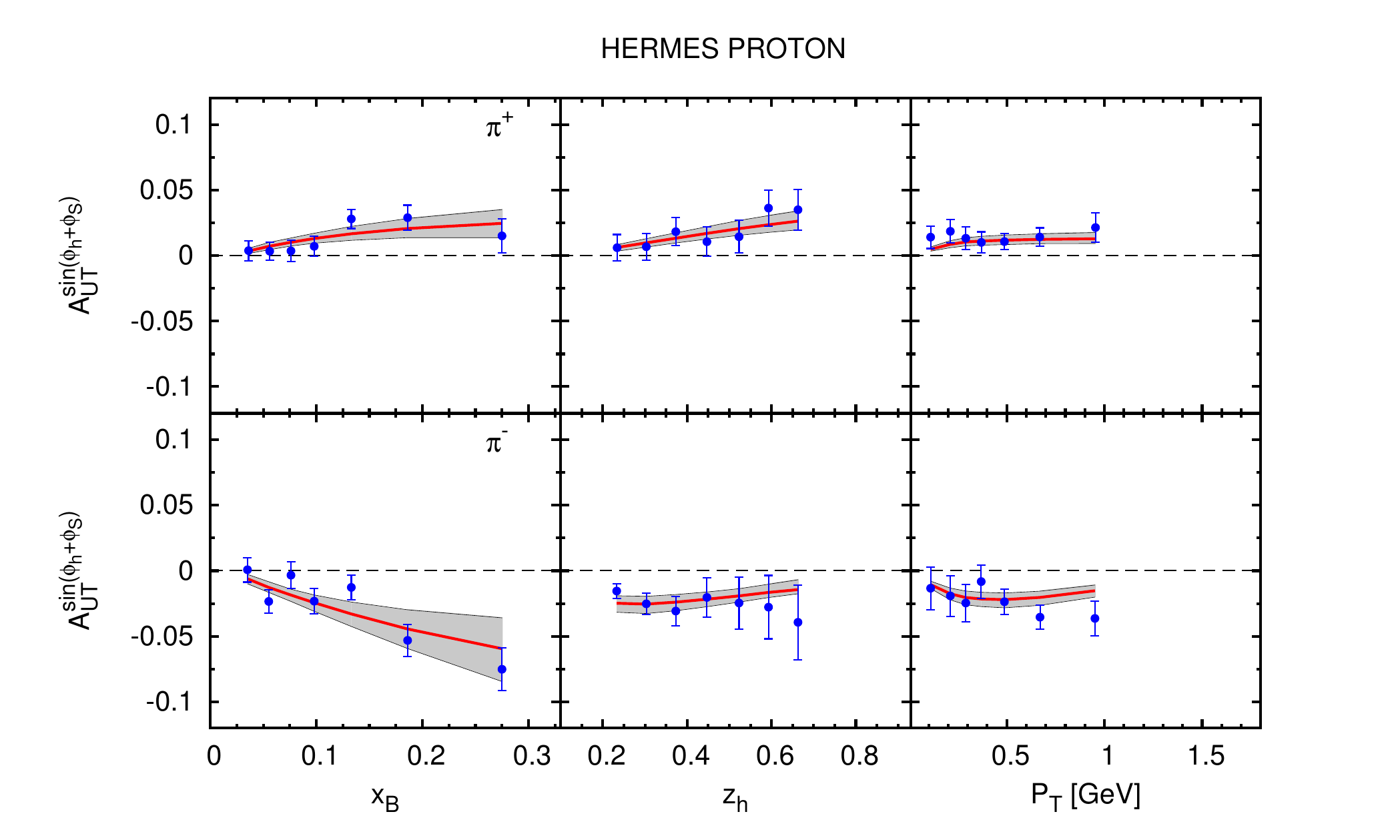}
\caption{The experimental data on the SIDIS azimuthal moment
$A_{UT}^{\sin(\phi_h+\phi_S)}$ as measured by the HERMES
Collaboration~\cite{Airapetian:2010ds}, are compared to the curves obtained from
our global reference fit. The solid lines correspond to the parameters given
in Table~\ref{tab:fitpar}, while the shaded areas correspond to the statistical 
uncertainty on these parameters, as explained in the text. Notice that, at 
order $k_\perp/Q$ and $p_\perp/Q$, $x_B=x$ and $z_h=z$.}
\label{fig:an-hermes-pip}
\end{figure}

\begin{figure}[h!t]
\includegraphics[width=9.9truecm,angle=0]{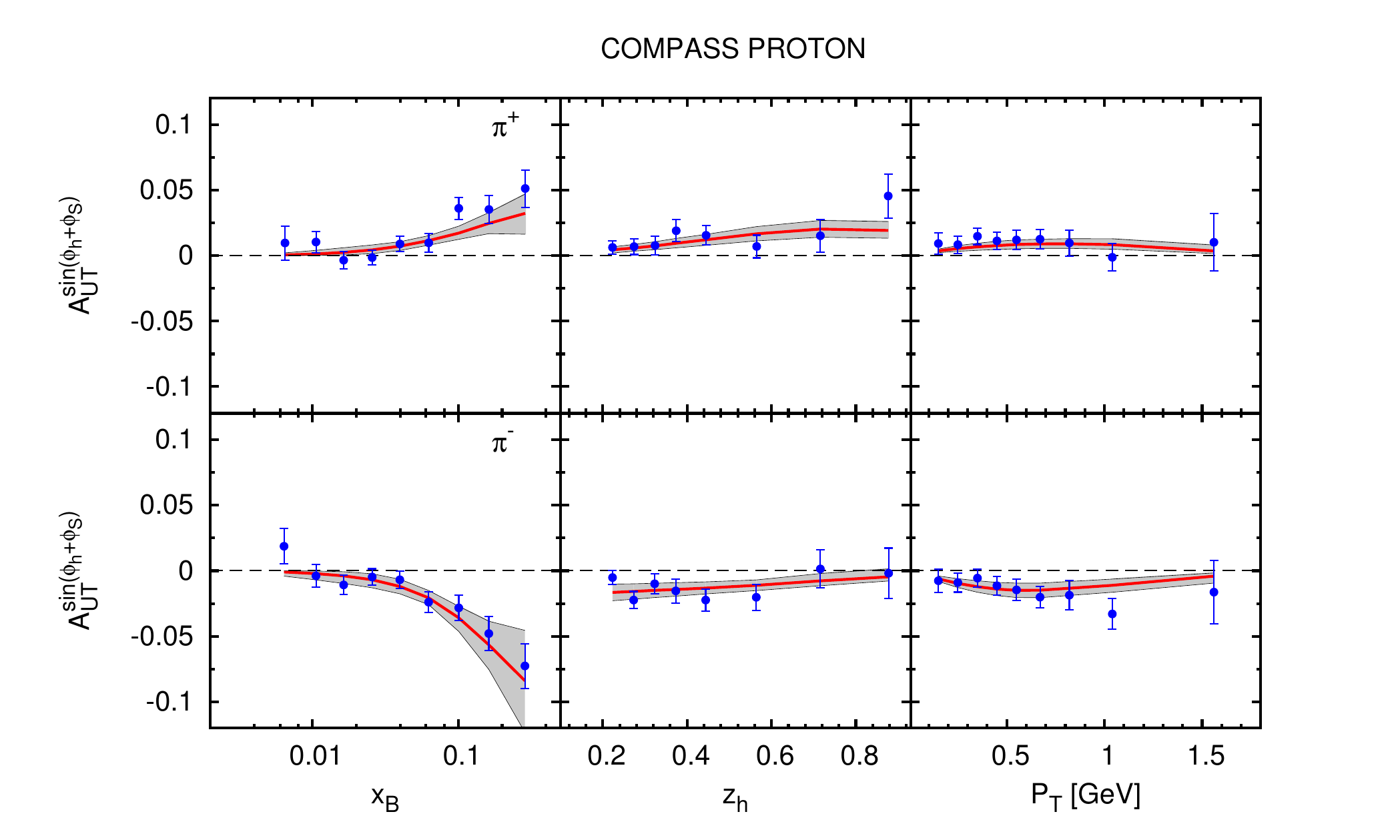}\\
\includegraphics[width=9.9truecm,angle=0]{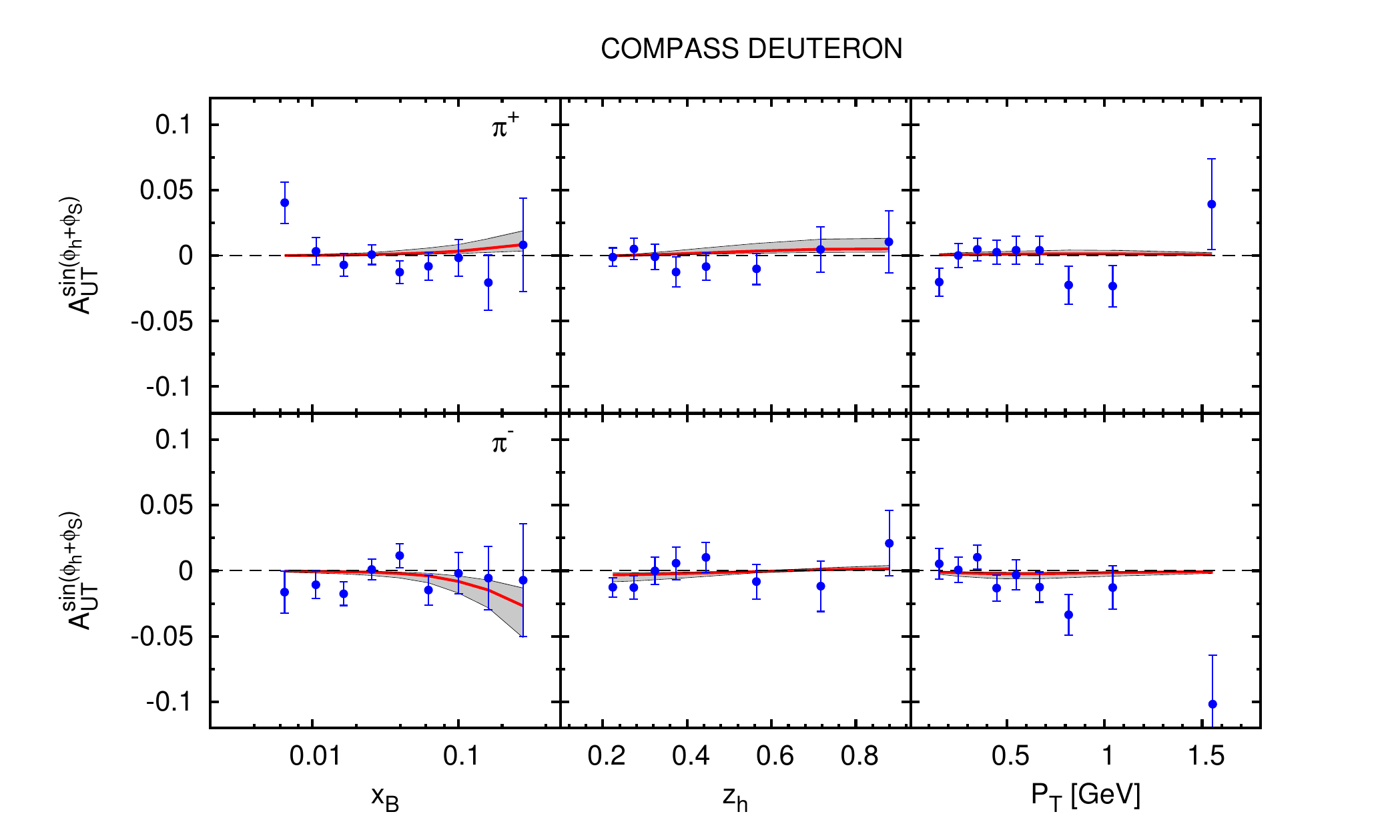}
\caption{The experimental data on the SIDIS azimuthal moment
$A_{UT}^{\sin(\phi_h+\phi_S)}$ as measured by the COMPASS Collaboration on
proton (upper panel) and deuteron (lower panel) targets~\cite{Martin:2013eja,Adolph:2014zba},
are compared to the curves obtained from our global reference fit.
The solid lines correspond to the parameters given in Table~\ref{tab:fitpar}, while
the shaded areas correspond to the statistical uncertainty on these parameters,
as explained in the text. Notice that, at  
order $k_\perp/Q$ and $p_\perp/Q$, $x_B=x$ and $z_h=z$.
}
\label{fig:an-compass-pip}
\end{figure}

\begin{figure}[h!t]
\includegraphics[width=6.5truecm,angle=0]{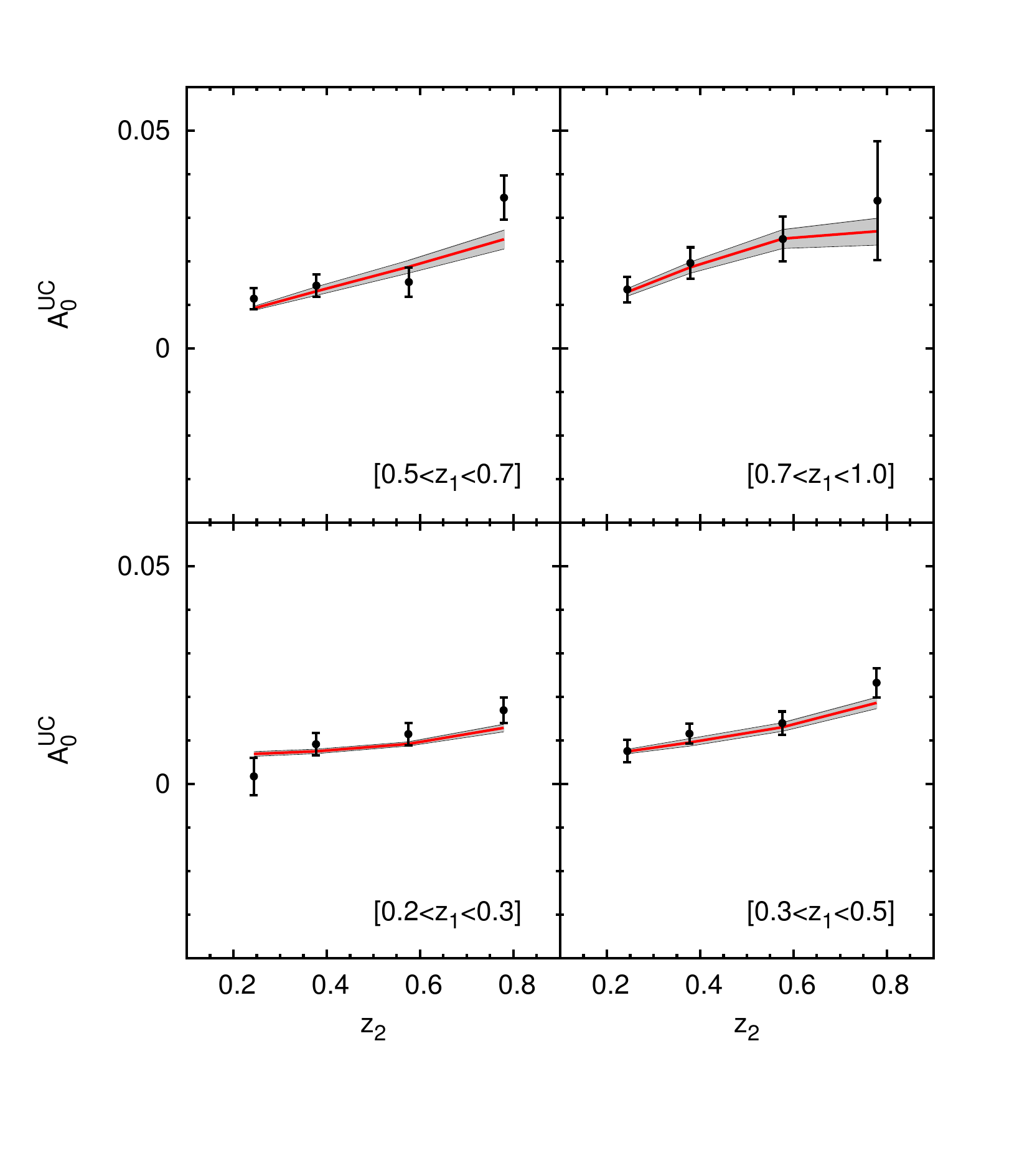}
\includegraphics[width=6.5truecm,angle=0]{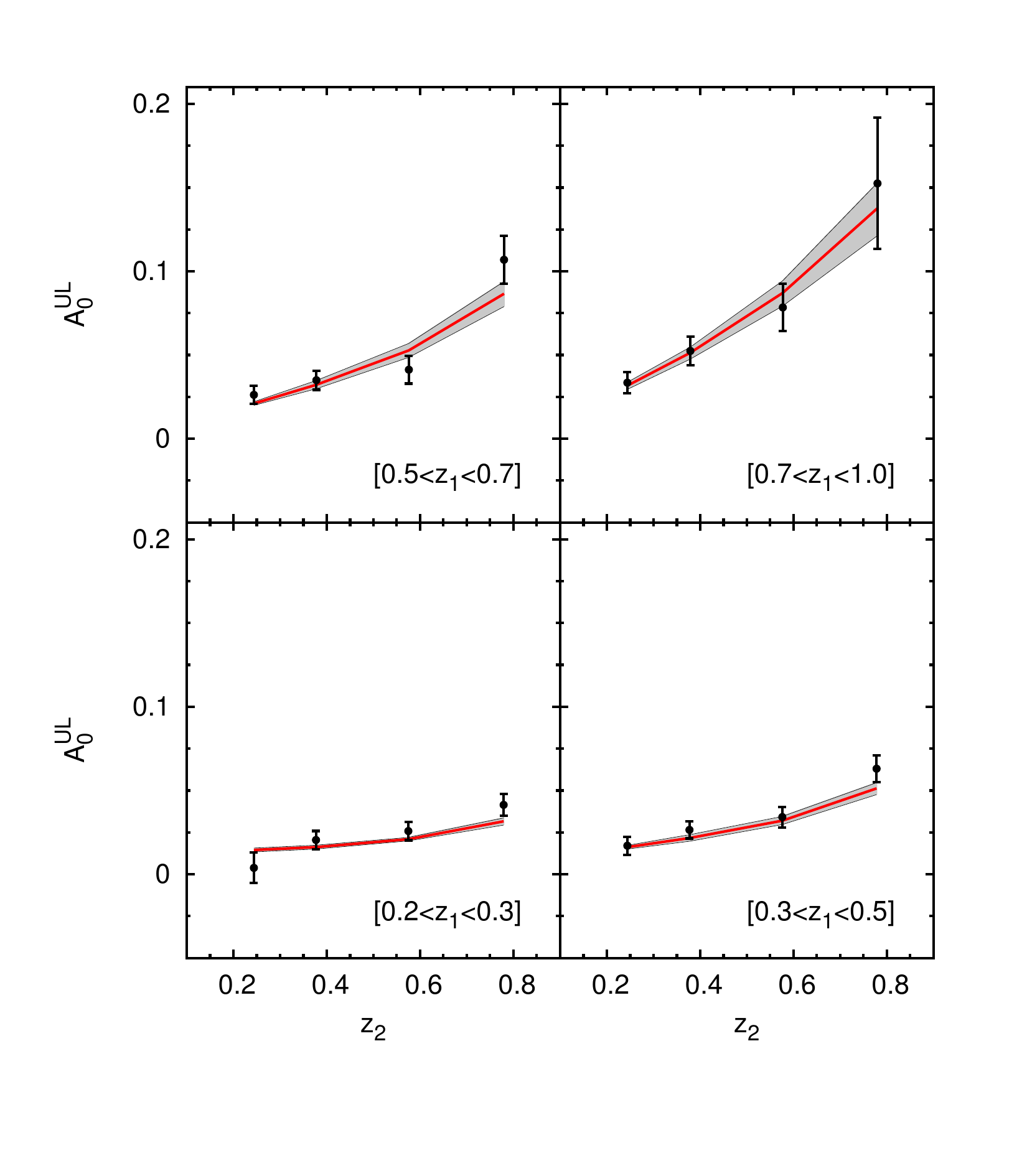}
\caption{The experimental data on the azimuthal correlations
$A_{0}^{UC}$ (left panel) and $A_{0}^{UL}$ (right panel) as functions of $z_1$ and $z_2$ in unpolarised
$e^+e^- \to h_1 \, h_2 \, X$ processes, as measured by the Belle
Collaboration~\cite{Seidl:2008xc,Seidl:2012er}, are compared to the curves obtained from
our global reference fit. The solid lines correspond to the parameters given
in Table~\ref{tab:fitpar}, while the shaded areas correspond to the statistical
uncertainty on these parameters, as explained in the text.
}
\label{fig:an-belle-pip}
\end{figure}

\begin{figure}[h!t]
\includegraphics[width=10.0truecm,angle=0]{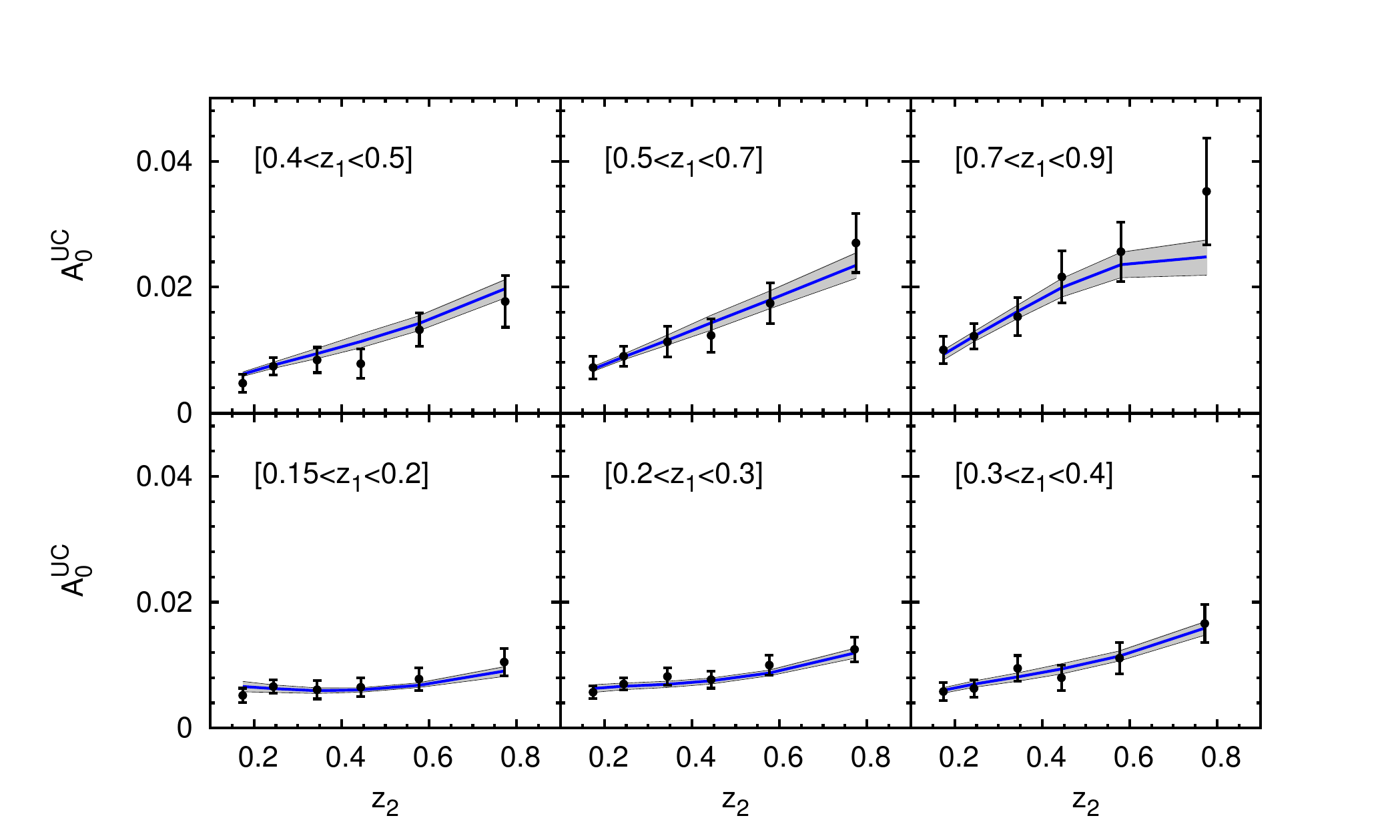}
\includegraphics[width=10.0truecm,angle=0]{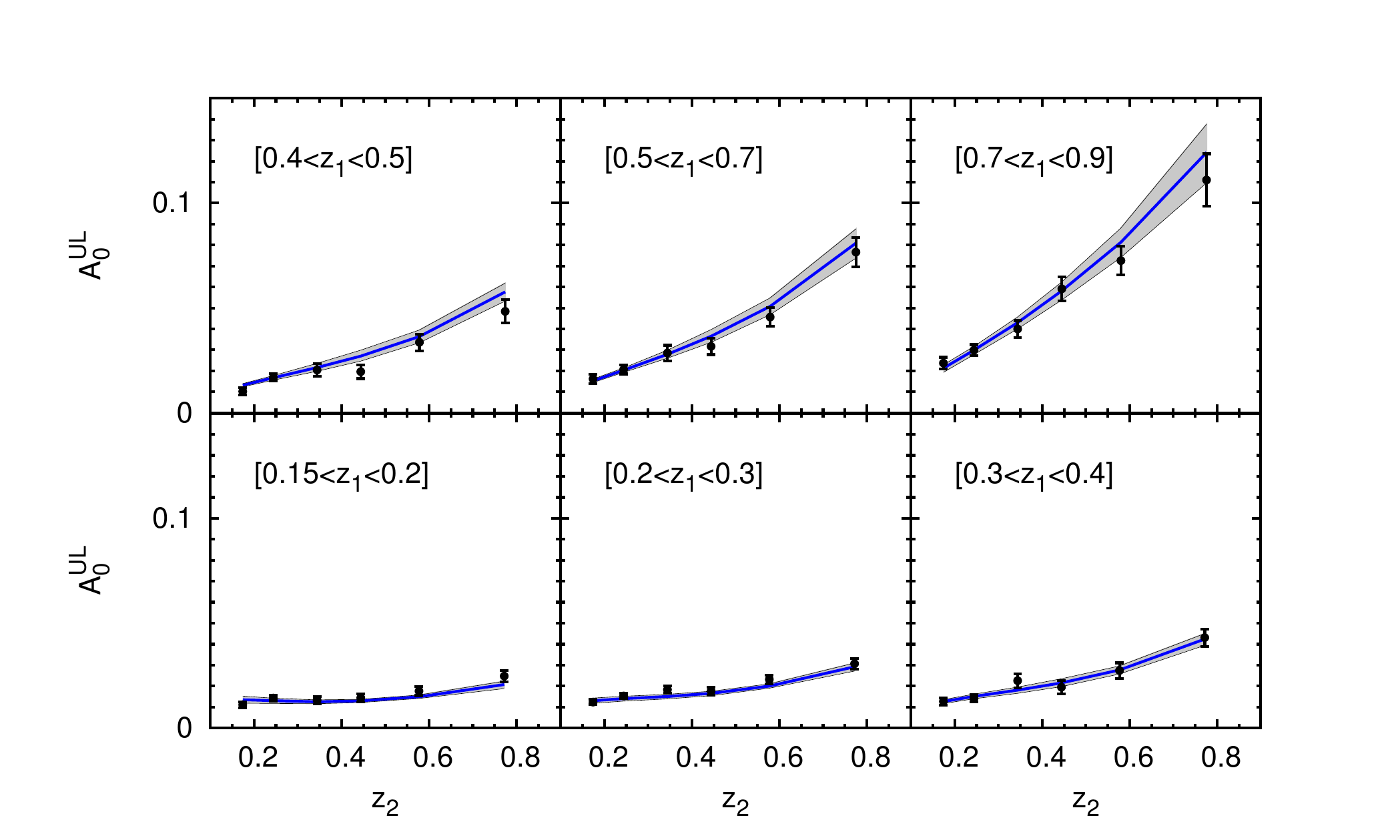}
\caption{The experimental data on the azimuthal correlations
$A_{0}^{UC}$ (upper panel) and $A_{0}^{UL}$ (lower panel) as functions of $z_1$ and $z_2$ in unpolarised
$e^+e^- \to h_1 \, h_2 \, X$ processes, as measured by the BaBar
Collaboration~\cite{TheBABAR:2013yha}, are compared to the curves obtained
from our global reference fit. The solid lines correspond to the parameters
given in Table~\ref{tab:fitpar}, while the shaded areas correspond to the statistical
uncertainty on these parameters, as explained in the text.}
\label{fig:an-babar-pip}
\end{figure}

\begin{figure}[h!t]
\includegraphics[width=7.truecm,angle=0]{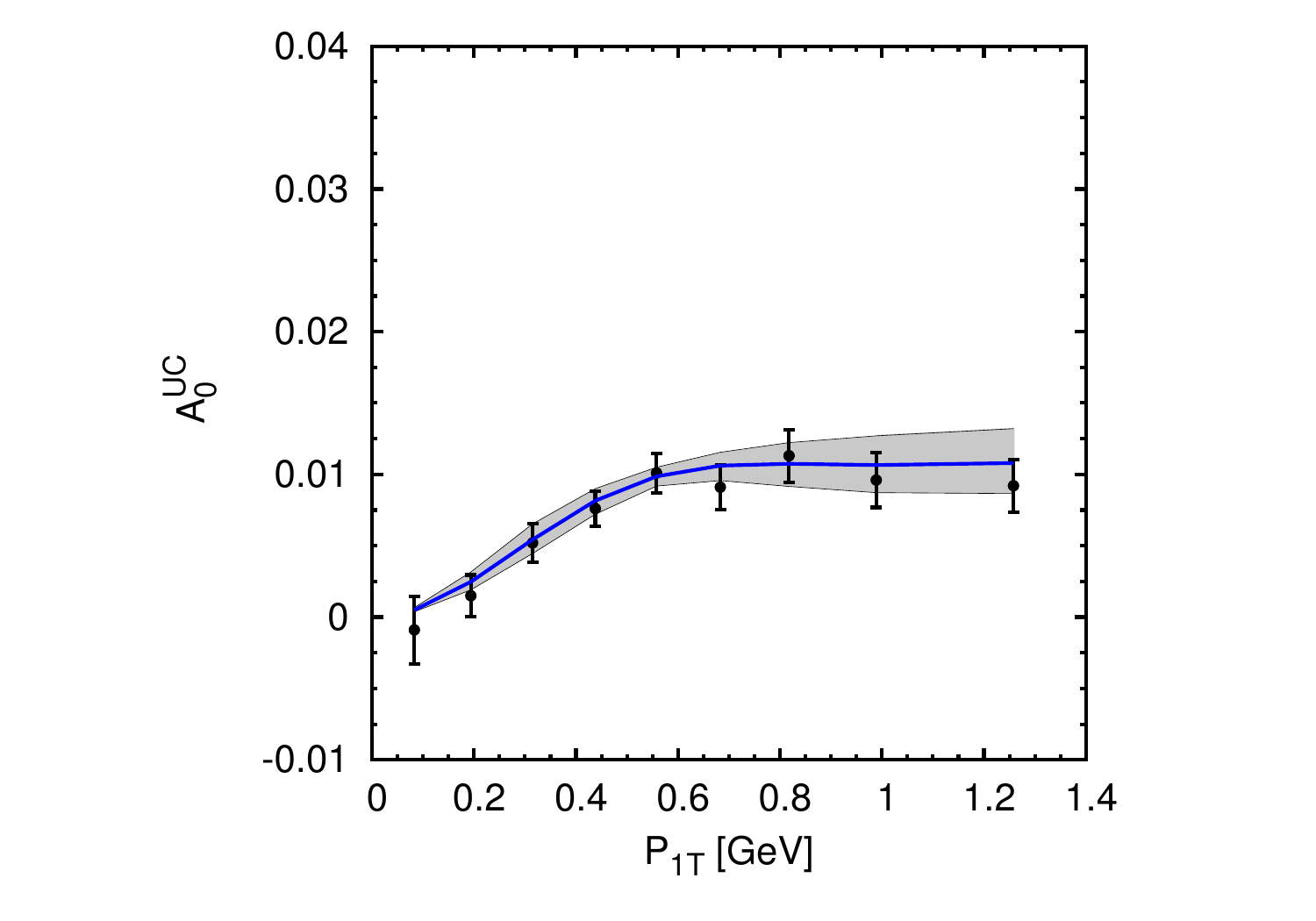}
\includegraphics[width=7.truecm,angle=0]{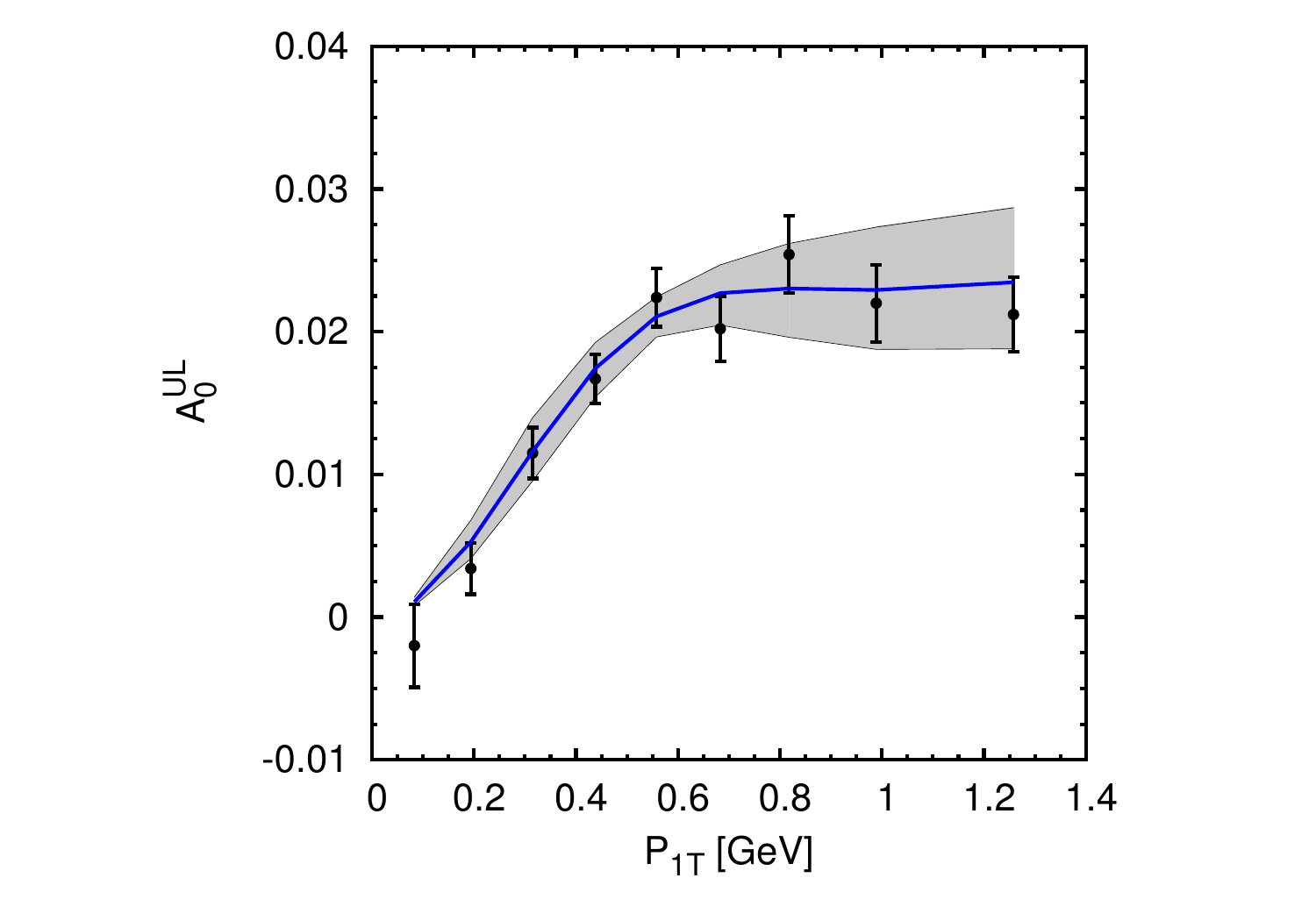}
\caption{The experimental data on the azimuthal correlations
$A_{0}^{UC}$ (left panel) and $A_{0}^{UL}$ (right panel) as functions of $P_{1T}$ in unpolarised
$e^+e^- \to h_1 \, h_2 \, X$ processes, as measured by the BaBar
Collaboration~\cite{TheBABAR:2013yha}, are compared to the curves obtained from
our global reference fit. The solid lines correspond to the parameters given
in Table~\ref{tab:fitpar}, while the shaded areas correspond to the statistical
uncertainty on these parameters, as explained in the text.}
\label{fig:an-babar-pt0}
\end{figure}

\begin{figure}[h!t]
\includegraphics[width=5.2truecm,angle=0]{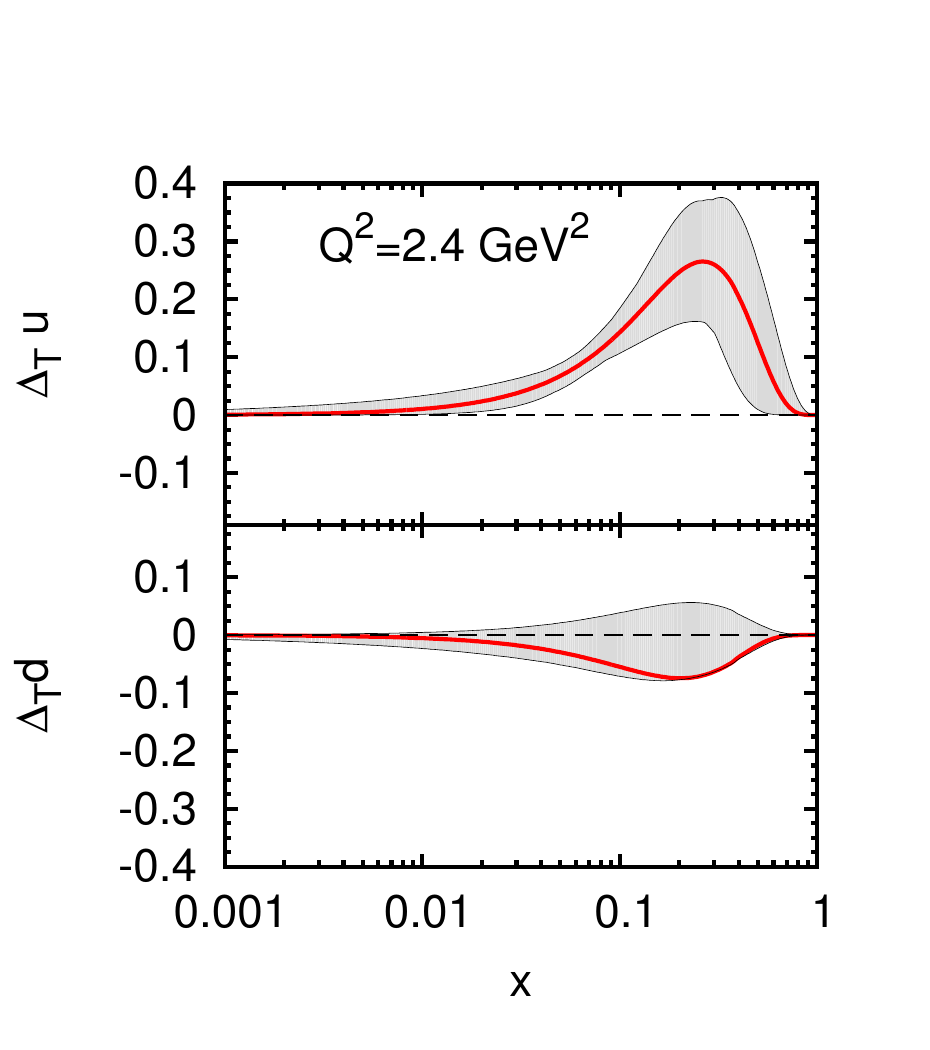}
\includegraphics[width=10.truecm,angle=0]{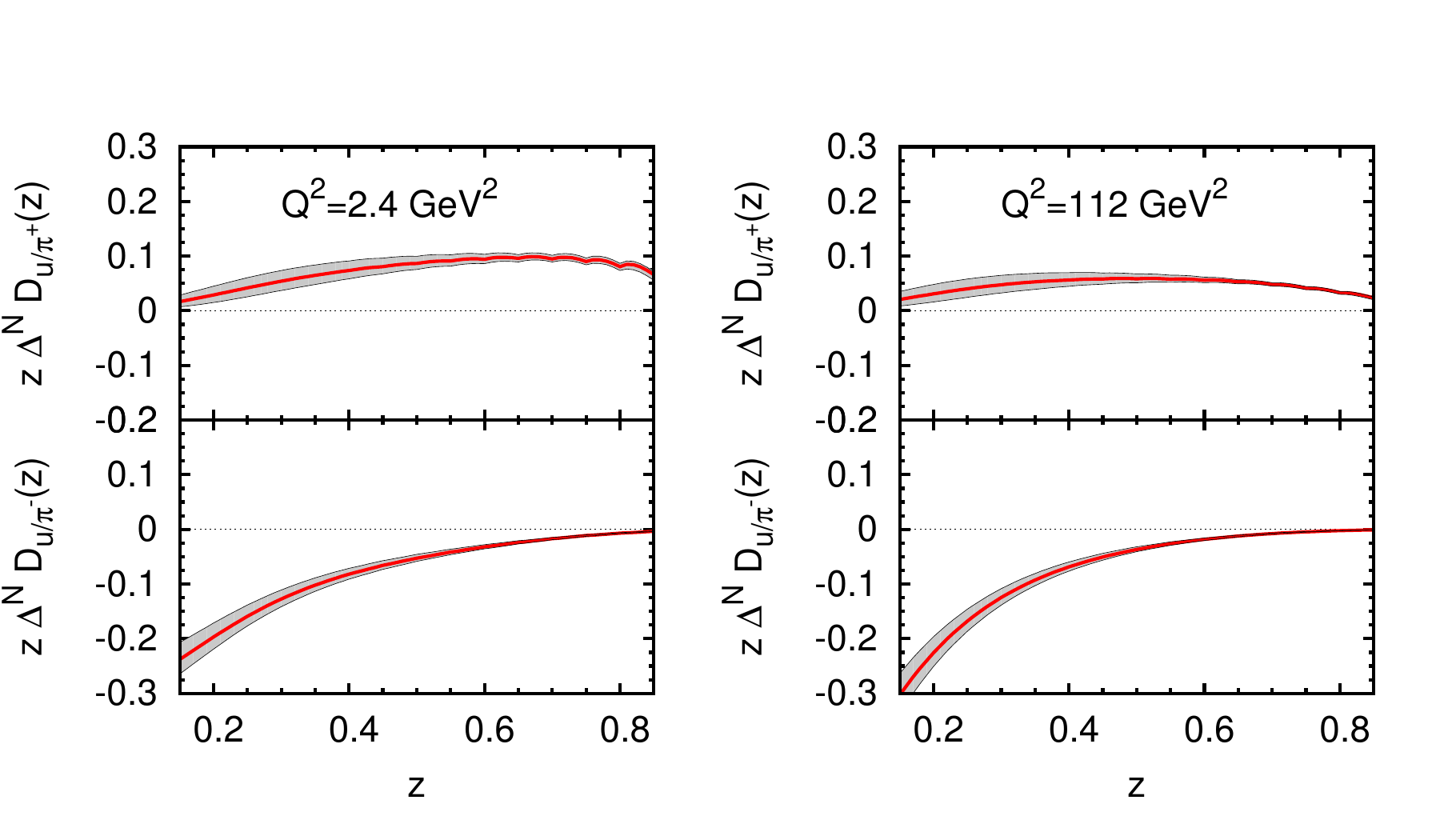}\\
\caption{Our best fit results for the valence $u$ and $d$ quark transversity
distributions at $Q^2 = 2.4$ GeV$^2$ (left panel) and for the lowest $\bfp_\perp$ moment of 
the favoured and disfavoured Collins functions at $Q^2 = 2.4$ GeV$^2$ (central panel) and at
$Q^2=112$ GeV$^2$ (right panel). The solid lines correspond to the parameters
given in Table~\ref{tab:fitpar}, while the shaded areas correspond to the statistical uncertainty 
on 
these parameters, as explained in the text.
}
\label{fig:transv-coll}
\end{figure}

\begin{figure}[h!t]
\includegraphics[width=5.5truecm,angle=0]{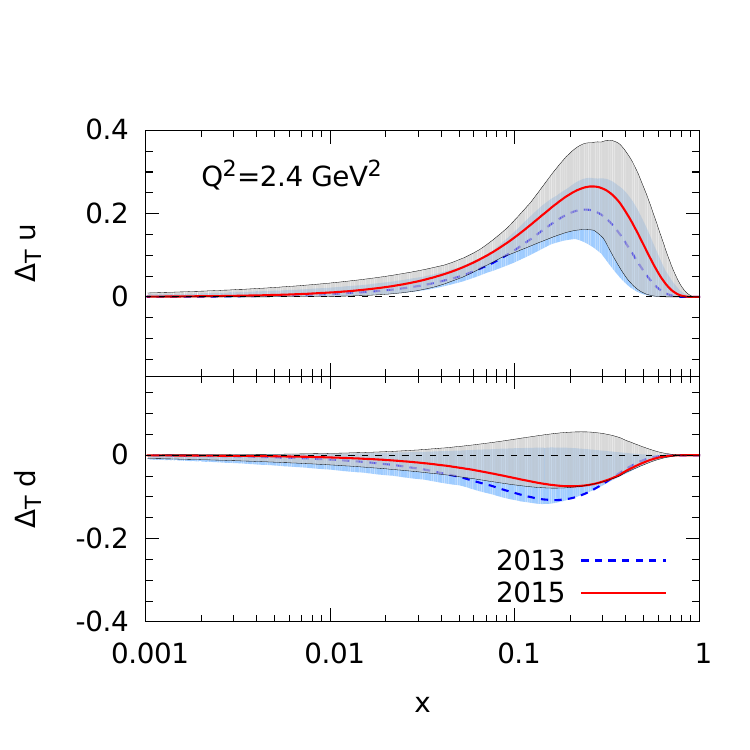}
\hspace*{0.5cm}
\includegraphics[width=5.5truecm,angle=0]{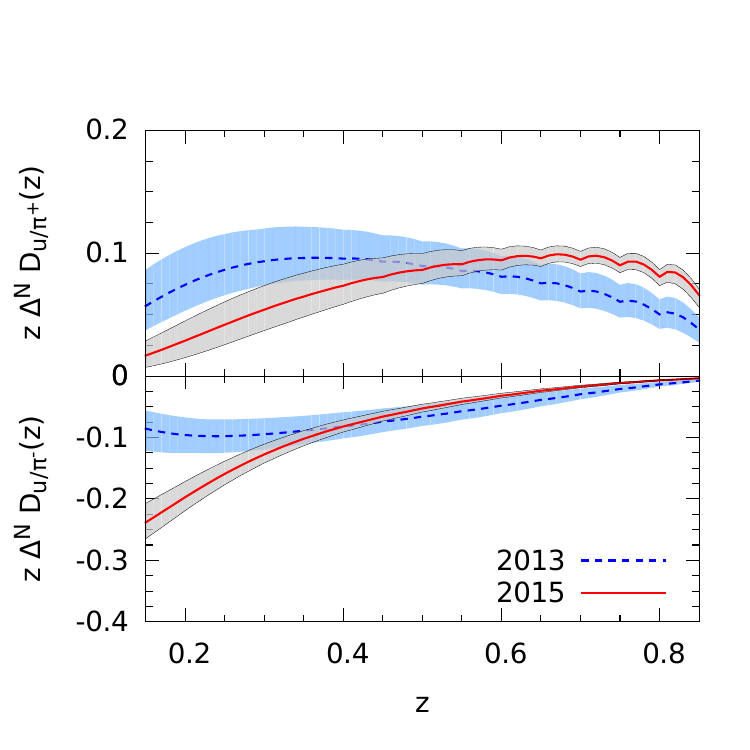}
\caption{Comparison of our reference best fit results (red, solid lines)
for the valence $u$ and $d$ quark transversity distributions (left panel)
and for the lowest $\bfp_\perp$ moment of the favoured and disfavoured Collins functions (right panel),
at $Q^2 = 2.4$ GeV$^2$,  with those from our previous
analysis~\cite{Anselmino:2013vqa} (blue, dashed lines).
}
\label{fig:transv-coll-comparison}
\end{figure}

\begin{figure}[h!t]
\includegraphics[width=5.5truecm,angle=0]{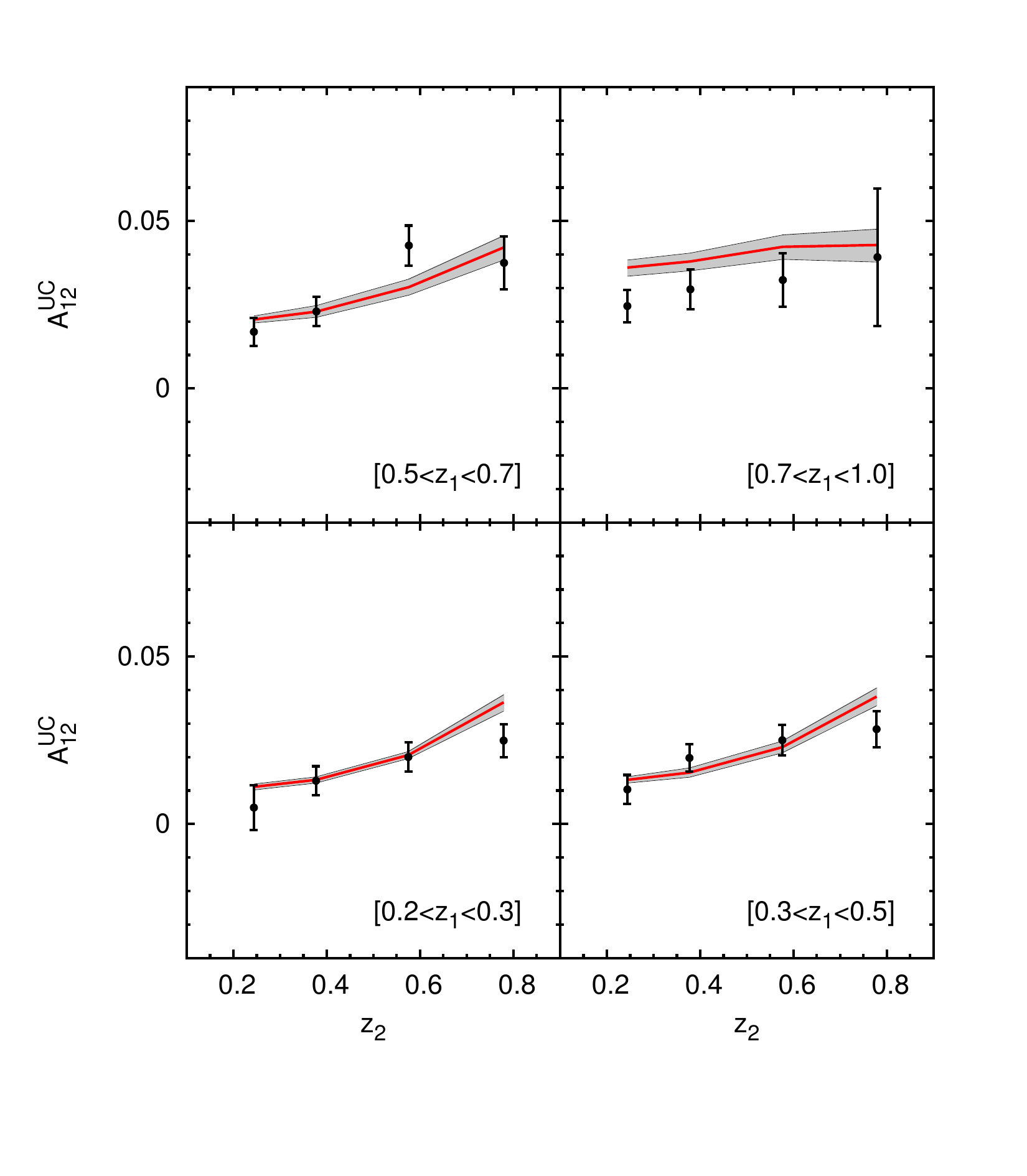}
\includegraphics[width=5.5truecm,angle=0]{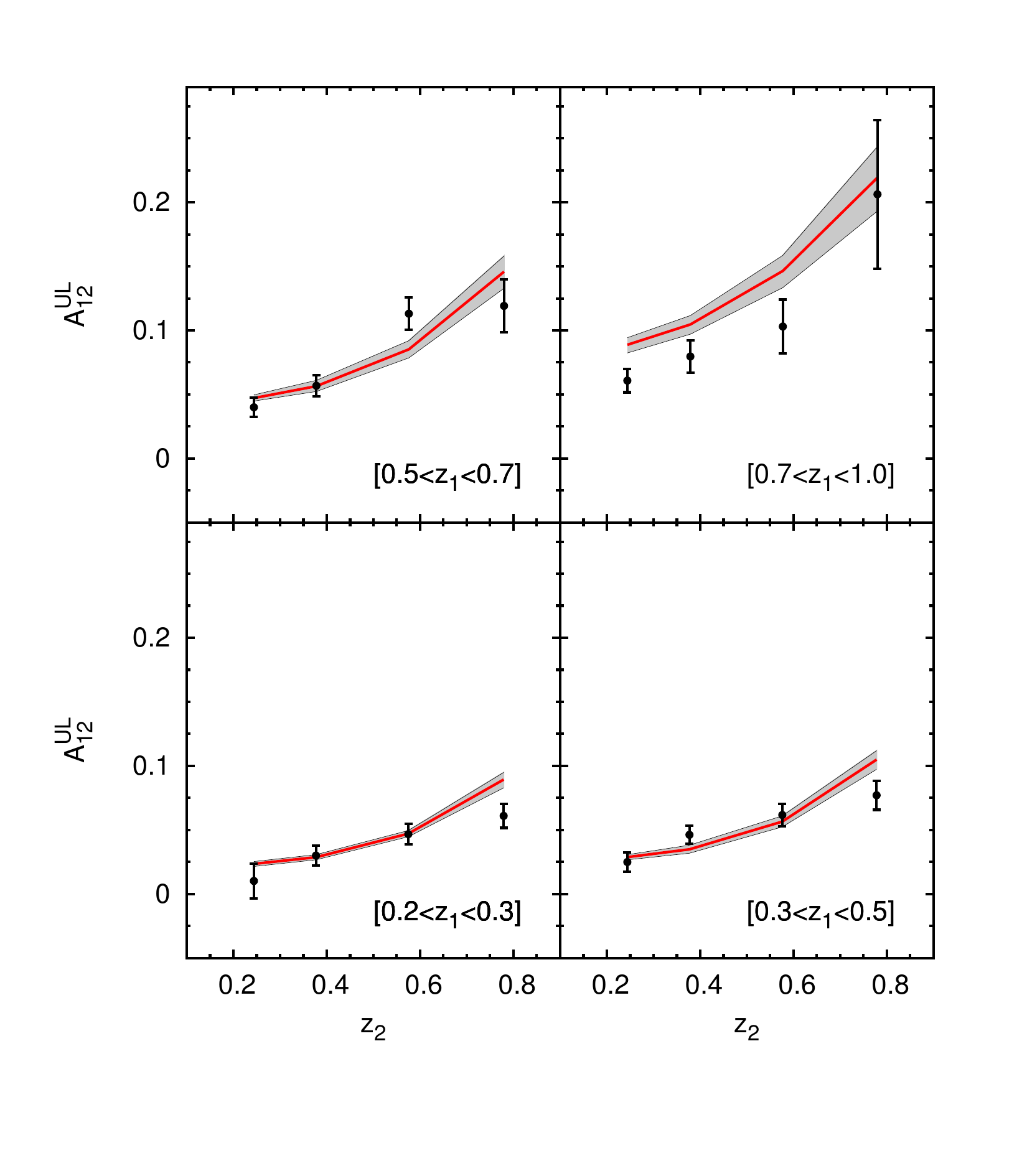}
\caption{The experimental data on the azimuthal correlations
$A_{12}^{UC}$ (left panel) and $A_{12}^{UL}$ (right panel) as functions of $z_1$ and $z_2$ in unpolarised
$e^+e^- \to h_1 \, h_2 \, X$ processes, as measured by the Belle
Collaboration~\cite{Seidl:2008xc,Seidl:2012er}, are compared to the curves given by the
parameters shown in Table~\ref{tab:fitpar}. The shaded areas correspond to the
statistical uncertainty on these parameters, as explained in the text.
These data have not been used in the global reference fit.}
\label{fig:an-belle-12-pip}
\end{figure}

\begin{figure}[h!t]
\includegraphics[width=10.0truecm,angle=0]{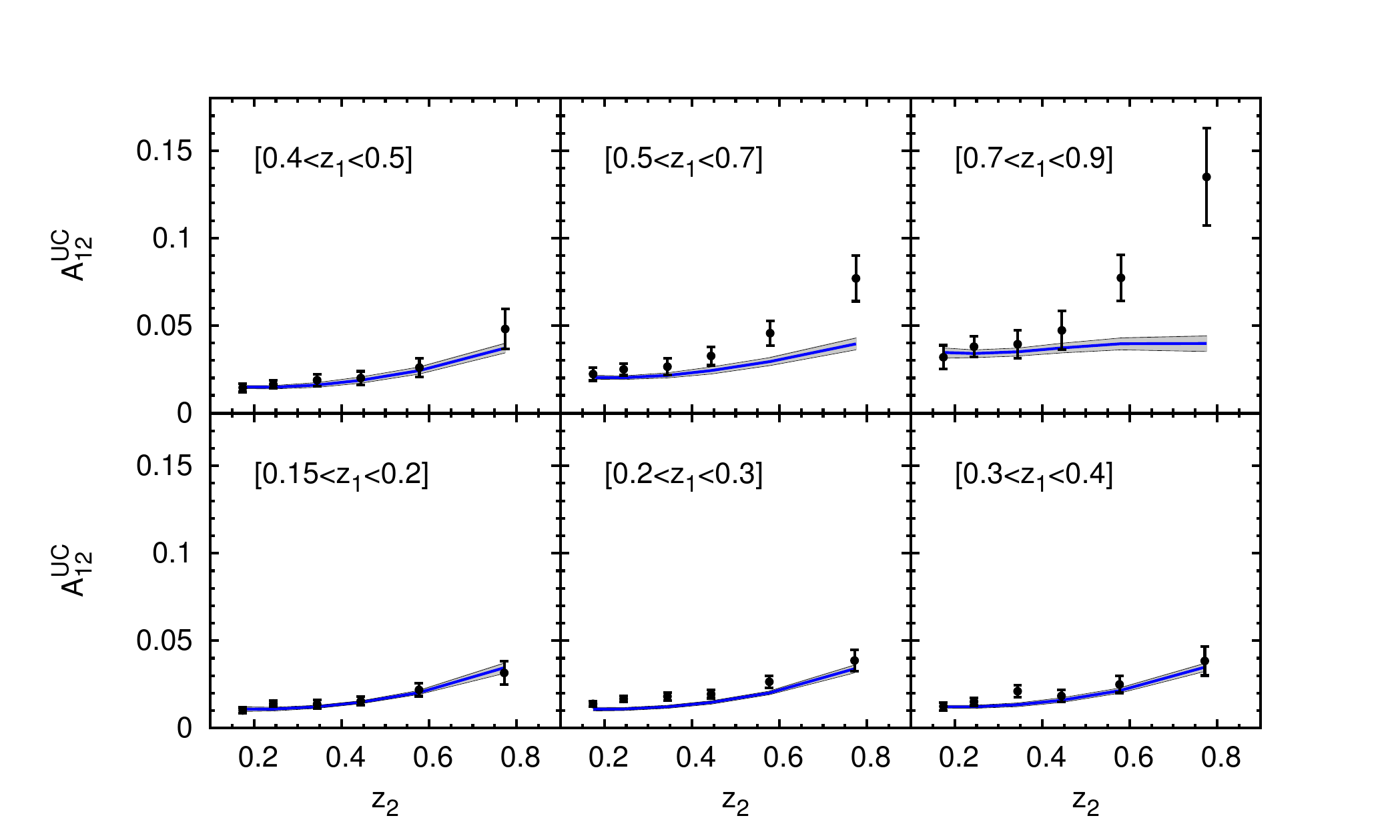}
\includegraphics[width=10.0truecm,angle=0]{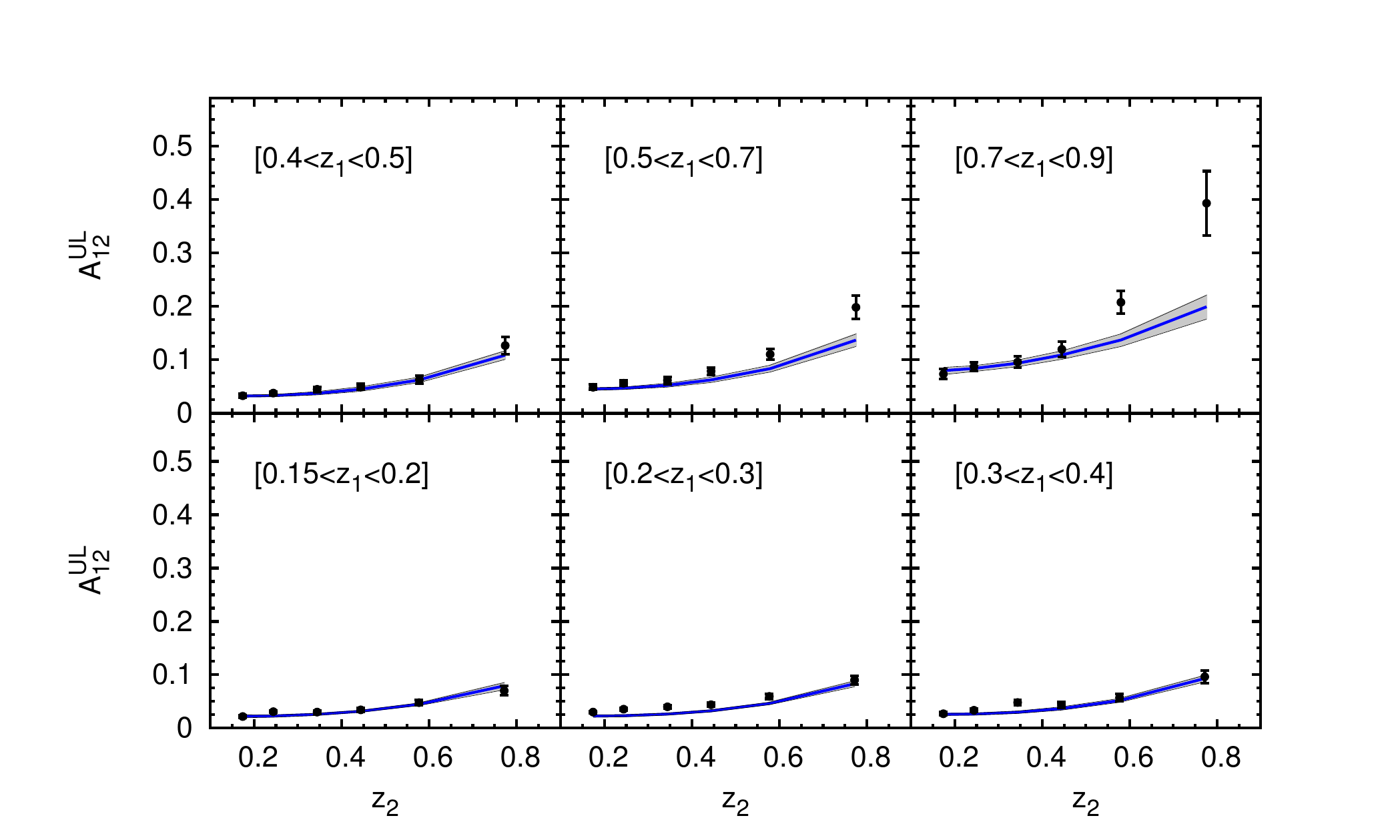}
\caption{The experimental data on the azimuthal correlations
$A_{12}^{UC}$ (upper panel) and $A_{12}^{UL}$ (lower panel) as functions of $z_1$ and $z_2$ in unpolarised
$e^+e^- \to h_1 \, h_2 \, X$ processes, as measured by the BaBar
Collaboration~\cite{TheBABAR:2013yha} are compared to the curves given by
the parameters shown in Table~\ref{tab:fitpar}. The shaded areas correspond to
the statistical uncertainty on these parameters, as explained in the text.
These data have not been used in the global reference fit.}
\label{fig:an-babar-12-pip}
\end{figure}

\begin{figure}[h!t]
\includegraphics[width=15.0truecm,angle=0]{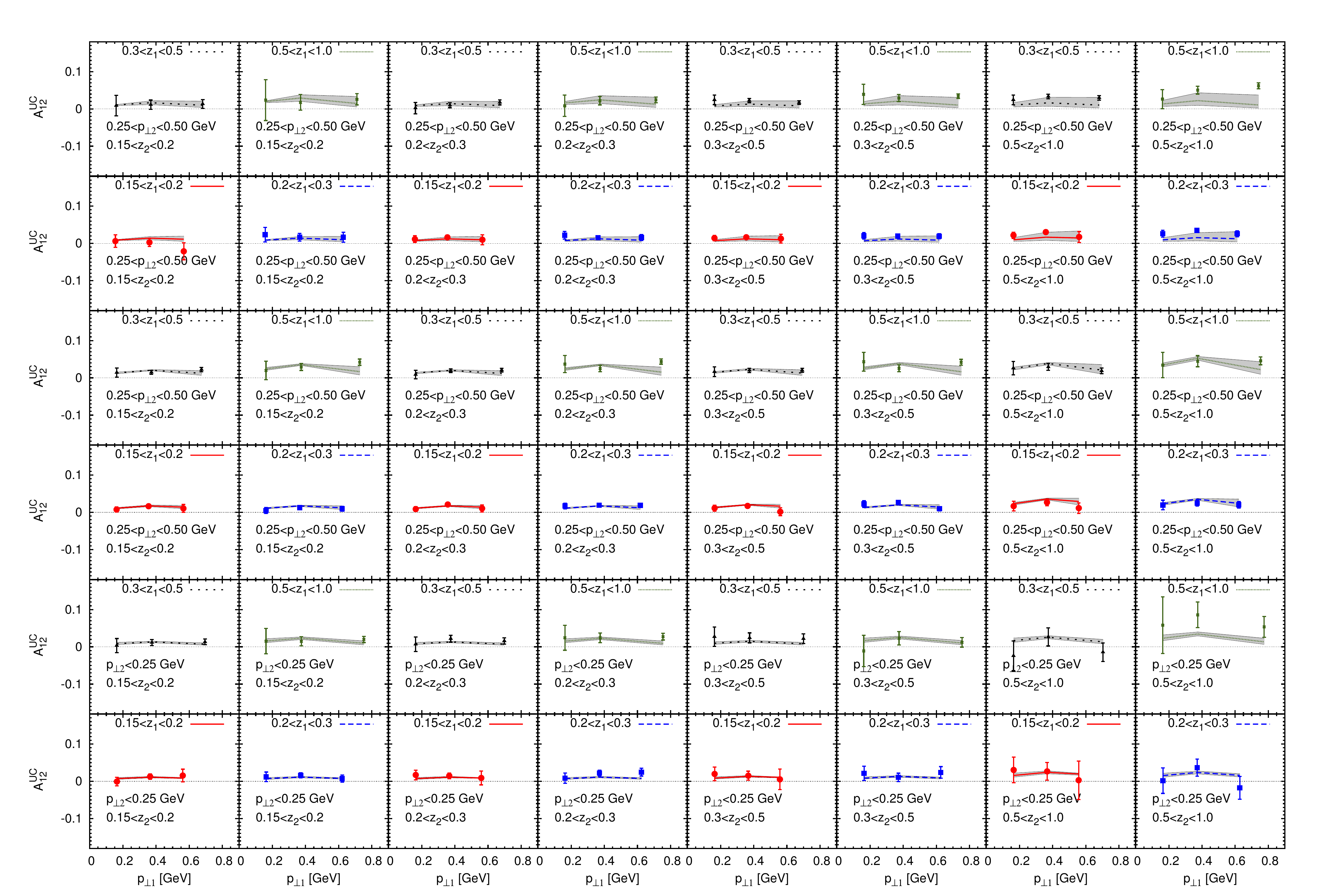}\\ \vspace*{1cm}
\includegraphics[width=15.0truecm,angle=0]{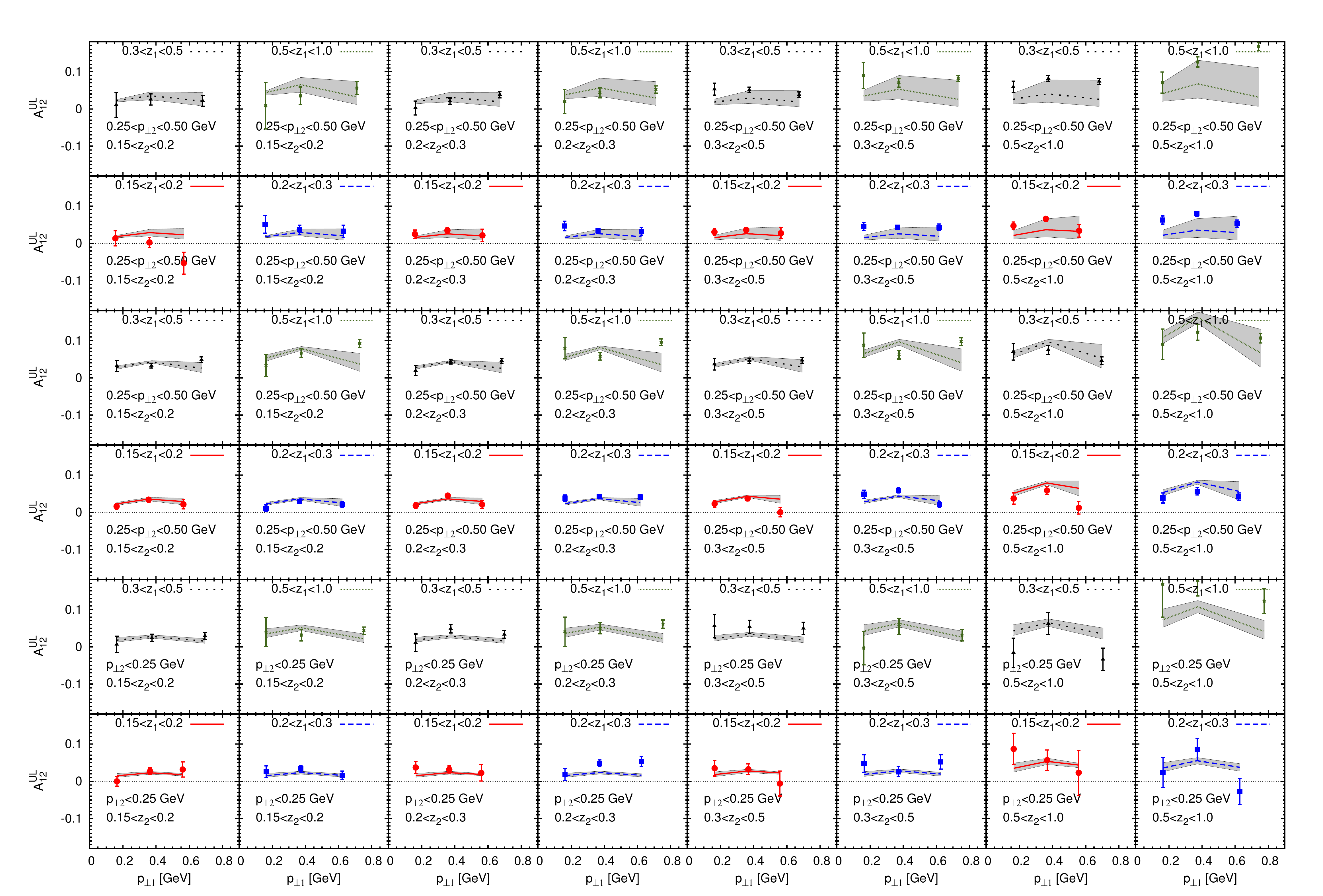}
\caption{The experimental data on the multidimensional azimuthal correlations
$A_{12}^{UC}$ (upper panel) and $A_{12}^{UL}$ (lower panel) in unpolarised $e^+e^- \to h_1 \, h_2 \, X$
processes, as measured by the BaBar Collaboration~\cite{TheBABAR:2013yha},
are compared to the curves given by the parameters shown in Table~\ref{tab:fitpar}.
The shaded areas correspond to the statistical uncertainty on these parameters,
as explained in the text. These data have not been used in the global
reference fit.}
\label{fig:an-babar-multi-pip}
\end{figure}

Table~\ref{tab:fitpar} reports the values of the parameters as determined by
the best fitting procedure, while in Table~\ref{tab:chisq} we summarise the
total $\chi^2$s of the fit and the $\chi^2$ contributions corresponding
to SIDIS and $e^+e^-$ experiments separately. As one can see, this fit is
very good. All data sets are very well reproduced,
as shown in Figs.~\ref{fig:an-hermes-pip}--\ref{fig:an-babar-pt0}.
The statistical errors shown in Table~\ref{tab:fitpar} and the bands in
Figs.~\ref{fig:an-hermes-pip}--\ref{fig:besIII-pt0-evot}
are obtained by sampling 1850 sets of parameters corresponding to a $\chi^2$ value in the
range between $\chi^2_{\rm min}$ and $\chi^2_{\rm min} + \Delta \chi^2 $,
as explained in Ref.~\cite{Anselmino:2008sga}. The value of $\Delta \chi^2$
corresponds to 95.45\% confidence level for 9 parameters;
in this case we have $\Delta \chi^2=17.2\,$.
%
%
%
\begin{table}[t]
\renewcommand{\tabcolsep}{0.4pc} 
\renewcommand{\arraystretch}{1.5} 
\begin{tabular}{|l|l|}
\hline
 $ N_{u_v}^{T} =0.61^{+0.39}_{-0.23} $ & $N_{d_v}^{T} =-1.00^{+1.86}_{-0.00}$ \\
 $ \alpha =0.70^{+1.31}_{-0.63} $ & $\beta =1.80^{+7.60}_{-1.80}$ \\
\hline
 $ N_{\rm fav}^{C} =0.90^{+0.09}_{-0.34} $ & $N_{\rm dis}^{C} =-0.37^{+0.05}_{-0.05}$ \\
 $ \gamma =2.02^{+0.83}_{-0.33} $ & $\delta =0.00^{+0.42}_{-0.00}$ \\
 $ M^2_C  =0.28^{+0.20}_{-0.09} \textrm{ GeV}^2 $ &  \\
\hline
\end{tabular}
\caption{
Best reference fit values of the 9 free parameters fixing the $u$ and $d$ valence quark
transversity distribution functions and the favoured and
disfavoured Collins fragmentation functions, as obtained by fitting
simultaneously HERMES and COMPASS data on the Collins asymmetry and
Belle and BaBar data on $A_{0}^{UL}$ and $A_{0}^{UC}$.}
\label{tab:fitpar}
\end{table}
%
\begin{table}[b]
\setlength{\arraycolsep}{50pt}
\renewcommand{\tabcolsep}{0.4pc} 
\renewcommand{\arraystretch}{1.5} 
\begin{tabular}{|c|c|c|c|}
\hline
 Experiment    &$\chi^2$ & n. points & $\chi^2/{\rm points}$ \\
\hline
$  \BELLEZ \quad A_0^{UL} $&  $ 14.0$ &$ 16$ & $0.88$\\
$  \BELLEZ \quad A_0^{UC} $&  $ 13.6$ &$ 16$ & $0.85$\\
$  \BABARZ \quad A_0^{UL} $&  $ 37.3$ &$ 36$ & $1.04$\\
$  \BABARZ \quad A_0^{UC} $&  $ 13.0$ &$ 36$ & $0.36$\\
$  \BABARPTZ \quad A_0^{UL} $&  $ 5.6$ &$ 9$ & $0.63$\\
$  \BABARPTZ \quad A_0^{UC} $&  $ 3.1$ &$ 9$ & $0.35$\\
\hline
Total \quad $ A_0 $&  $ 86.7$ &$ 122$ & $0.71$\\
\hline
\hline
HERMES p & 31.6 & 42 &  0.75\\
COMPASS p & 40.2 & 52 &  0.77\\
COMPASS d & 58.5 & 52 &  1.12\\
\hline
Total SIDIS& 130.3 & 146 &  0.89\\
\hline
\hline
Total  & 217.0 & 268 & $\chi^2_{\rm d.o.f.}=0.84$ \\
\hline
\end{tabular}
\caption{Contributions of each individual set of fitted data to the total
$\chi ^2$ of our reference fit. The upper part of the table refers to
$e^+e^-$ data. Here we show the $\chi^2$s obtained for the Belle and
BaBar $A_0^{UL}$ and $A_0^{UC}$ asymmetries as functions of $z_1$ and $z_2$  (integrated
over the hadronic transverse momentum $P_{1T}$) and as a functions of $P_{1T}$
(integrated over $z_1$ and $z_2$). The second part refers to SIDIS measurements
off proton and deuteron targets. In the last line we report the total $\chi^2$
and $\chi^2_{\rm d.o.f.}$ of the fit.}
\label{tab:chisq}
\end{table}

Figs.~\ref{fig:an-hermes-pip} and~\ref{fig:an-compass-pip} show our best
fit results for the azimuthal modulation $A_{UT}^{\sin(\phi_h+\phi_S)}$ as
measured by the HERMES~\cite{Airapetian:2010ds} and COMPASS~\cite{Martin:2013eja}
Collaborations in SIDIS processes, while Figs.~\ref{fig:an-belle-pip}
and~\ref{fig:an-babar-pip} show our description of the azimuthal correlations
$A_{0}^{UL}$ and $A_{0}^{UC}$, as functions of $z_1$ and $z_2$ in unpolarised
$e^+e^- \to h_1 h_2 \, X$ processes, measured by the Belle~\cite{Seidl:2008xc,Seidl:2012er}
and BaBar~\cite{TheBABAR:2013yha} Collaborations, respectively.
Fig.~\ref{fig:an-babar-pt0} shows our best fit of the BaBar $A_{0}^{UL}$
and $A_{0}^{UC}$ asymmetries as functions of $P_{1T}$ ($p_{t0}$ in the
notation used by the BaBar Collaboration). We stress that these measurements
offer the first direct insight of the dependence of the Collins function on
the parton intrinsic transverse momentum: in fact, our global fit now
delivers a more precise determination of the Gaussian width of the Collins
function (through the $M_C$ parameter, see Table~\ref{tab:fitpar}), which in our
previous fits was affected by a very large uncertainty.

%
%
%
%
%
%

In Fig.~\ref{fig:transv-coll} we show the valence quark transversity
distributions and the lowest $\bfp_\perp$-moment of the favoured and
disfavoured Collins functions as extracted from our reference fit, while in
Fig.~\ref{fig:transv-coll-comparison} we compare them with those extracted
in our previous analysis~\cite{Anselmino:2013vqa}.
Notice that, in the case of a factorised Gaussian shape,
the lowest $\bfp_\perp$-moment of the Collins function,
\begin{equation}
\Delta^N D_{h/q^\uparrow}(z,Q^2) = \int\,d^2\bfp_\perp\,
\Delta^N D_{h/q^\uparrow}(z,p_\perp,Q^2)\,,
\label{p-0mom-D}
\end{equation}
is related to the $z$-dependent part of the Collins function,
$\tilde{\Delta}^N D_{h/q^\uparrow}(z,Q^2)$, Eqs.~(\ref{coll-funct}),~(\ref{coll-D}) and (\ref{hpcollins}),
by the simple relation
\be
\Delta ^N D_{h/q^\ua}(z, Q^2) = \frac{\sqrt{\pi}}{2} \,
\frac{\avp_{_{\!C}}^{3/2}}{\avp} \, \frac{\sqrt{2e}}{M_C} \,
\tilde{\Delta} ^N D_{h/q^\ua}(z, Q^2)\,.
\label{D-mom-tilde}
\ee
In Figs.~\ref{fig:transv-coll},~\ref{fig:transv-coll-comparison}
and~\ref{fig:comparison} we plot $\Delta ^N D_{h/q^\ua}(z,Q^2)$ in order to
facilitate the comparison with the results
of Refs.~\cite{Anselmino:2007fs, Anselmino:2008jk,Anselmino:2013vqa}.
{}From Fig.~\ref{fig:transv-coll-comparison} we can see that only the Collins
functions differ significantly;
this is due to the different choice of parameterisation. In fact, given
the lower statistics of the available data at that time, in 2013 we imposed
that the favoured and disfavoured ${\cal N}^{C}(z)$ functions had the same
$z$-dependence and could differ only by a normalisation constant, while in
this paper, where we can count on a much higher statistics, they are left
uncorrelated, with the disfavoured function being simply a constant multiplied
by the unpolarised fragmentation function, see Eqs.~\eqref{std-fav}
and~\eqref{std-dis}. The $u_v$ and $d_v$ transversity functions, instead,
are well compatible with their $u$ and $d$ counterparts extracted in 2013.
Notice that the present data actually allow the extraction of the sole $u_v$
transversity function, due to the strong $u$ dominance in the SIDIS data.
We have checked that $\Delta_T d = 0$ is a possible solution (and it is
in fact included in our uncertainty bands). Moreover, for instance, one 
could consider a scenario with only $u$ and $\bar u$ quark contributions, 
without any $d$ transversity distribution, obtaining a best fit of comparable 
quality. This might have an important
impact in the attempt to determine the tensor charge.

As mentioned above, in our global fit we include only the experimental
$e^+e^-$ measurements taken in the hadronic-plane reference frame, that
is only the $A_{0}^{UL}$ and $A_{0}^{UC}$ asymmetries are used to
constrain the model parameters. However, once the free parameters
have been determined by the best fit procedure, we can compare the
predictions obtained from our model with the measurements of the $A_{12}^{UL}$
and $A_{12}^{UC}$ asymmetries performed in the thrust-axis reference frame.
Figs.~\ref{fig:an-belle-12-pip}--\ref{fig:an-babar-multi-pip} show this
comparison: the predicted asymmetries are in satisfactory agreement with
experimental data, even for the multidimensional azimuthal correlations
(in bins of $z_1$, $z_2$, $\ppo$ and $\ppt$); there are only some problems
with data points corresponding to large values of $z_1$ and $z_2$, but this
is a delicate region where exclusive channels might contribute.

%
%
%
%
%
%
%

\section{On the role of alternative parameterisations and the $Q^2$ evolution
of the Collins function\label{Collins-evo}}


In Section~\ref{Mod} we performed a best fit by adopting a simple
phenomenological $Q^2$ evolution for the Collins function: we assumed the
ratio $\tilde{\Delta}^N D(z,Q^2)/D(z,Q^2)$ to be constant in $Q^2$, with the
unpolarised fragmentation function $D(z,Q^2)$ evolving with a DGLAP kernel.
However, the Collins function can be shown to be related to the collinear
$H^{(3)}_{h/q}$ twist-three fragmentation function~\cite{Yuan:2009dw}, the
diagonal part of which evolves with a transversity kernel as the transversity
function. Therefore, it is interesting to apply this kind of evolution to
the Collins function and study  the consequences of such an evolution on
our best fit.

To this purpose, we assume the $z$-dependent part of the Collins distribution,
$\tilde{\Delta} ^N D_{h/q^\ua}$, to evolve with a transversity kernel,
similarly to what is done for the transversity function, as suggested in
Refs.~\cite{Yuan:2009dw,Kang:2010xv}.
The results we obtain show a slight deterioration of the fit quality,
with a global $\chi^2_{\rm d.o.f.}$ increasing from $0.84$ to $1.20$.
Although this is still an acceptable result, one may wonder whether this
is a genuine effect of the chosen evolution model or, rather, a byproduct
of the functional form adopted for the Collins function parameterisation.

We have therefore exploited a different parameterisation based on a polynomial
form. In principle, the polynomial could be of any order. We have started by
using an order zero polynomial, then increased it to order one and,
subsequently, to order two. In doing so, we have seen that the quality of
the fit improves remarkably when going from order zero to order one
(i.e. from 2 to 4 free parameters) but it stops improving
when further increasing to higher orders. We therefore choose a first
order polynomial form, which has the added advantage of depending on the
same number of free parameters as the standard parameterisation
of Eqs.~\eqref{std-fav} and~\eqref{std-dis}.

We consider generic combinations of fixed order Bernstein
polynomials (see, for example, Ref.~\cite{Farouki:2012}) as they offer a relatively 
straightforward way to keep track of the appropriate normalisation:
\be
\mathcal{N}^{\C}_{i}(z)=a_i P_{01}(z)+b_i P_{11}(z) \qquad
i={\rm fav,\,dis}
\label{poly}
\ee
where $P_{01}(z)=(1-z)$ and $P_{11}(z)=z$ are Bernstein polynomials of
order one. Notice that by constraining the four free parameters in such
a way that $-1\le a_i \le +1$ and $-1 \le b_i \le +1$, the Collins
function automatically fulfils its positivity bounds, as in the standard
parameterisation. The Collins function will be globally modelled as shown
in Eqs.~\eqref{coll-funct} and~\eqref{coll-D}, with
$\mathcal{N}_{\rm fav}^{\C}(z)$ and $\mathcal{N}_{\rm dis}^{\C}(z)$ as
given in Eq.~\eqref{poly}.

It turns out that with a transversity-like $Q^2$ evolution of the Collins
function coupled to this polynomial parameterisation, we can obtain best fit
results of  similar quality as we found for our reference fit, with
$\chi^2_{\rm d.o.f.} = 1.00$. Notice that, adopting the polynomial
parameterisation and the standard evolution of the Collins function,
one would obtain $\chi^2_{\rm d.o.f.}=0.92$ and no improvement would be 
achieved with respect to our reference fit.

In Table~\ref{tab:evos} we show the $\chi^2_{\rm d.o.f.}$ corresponding to
different choices of evolution and parameterisation for the Collins functions.
As it can be seen, all $\chi^2_{\rm d.o.f.}$ are rather close to $1$:
this suggests that the observables we are fitting exhibit a very mild $Q^2$
dependence. In fact, we have checked that a similar $\chi^2_{\rm d.o.f.}$
can be obtained by not including any $Q^2$ dependence at all in the PDFs
and FFs. One of the reasons our model works well is that it allows
for an approximate cancellation of the $Q^2$-dependence in the asymmetries.

\begin{figure}[h!t]
\includegraphics[width=10.0truecm,angle=0]
{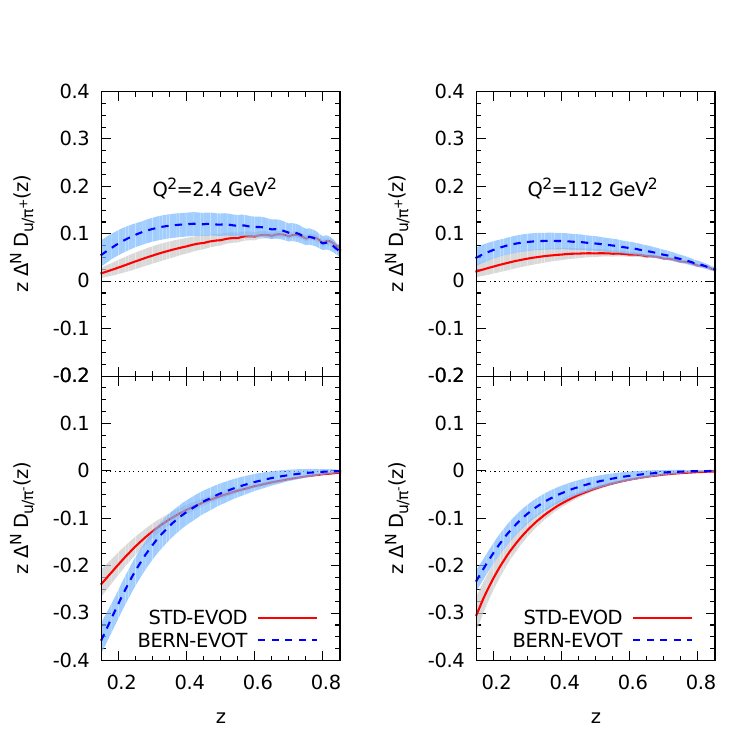}
\caption{Comparison of the lowest $\bfp_\perp$ moment, according to
Eq.~(\ref{p-0mom-D}) of the text, of the favoured (upper panels) and 
disfavoured (lower panels) Collins functions at $Q^2=2.4$ GeV$^2$ 
(left panel) and at $Q^2=112$ GeV$^2$ (right panel) obtained from best 
fit procedures using different evolution kernels and parameterisations. 
The solid red lines represent the Collins moments obtained by using the
standard parameterisation and employing the standard evolution. The dashed 
blue lines represent the same quantities obtained using the polynomial 
parameterisation and by evolving the Collins function with a transversity 
kernel. The shaded areas correspond to the statistical uncertainty on 
the best fit parameters, as explained in the text.}
\label{fig:comparison}
\end{figure}

Fig.~\ref{fig:comparison} shows a comparison between the Collins functions 
extracted from the same sets of data using the reference fit procedure (red, 
solid lines) and the transversity-like $Q^2$ evolution with a polynomial 
parameterisation (blue, dashed lines). 
No really significant differences can 
be noticed. We do not show the same comparison between the transversity 
distributions obtained in the two best fit procedures, as the differences 
would be hardly noticeable; this can be seen by comparing the values of 
the parameters $N_{u_v}^{\T}, N_{d_v}^{\T}, \alpha$ and $\beta$, fixing the 
transversity distributions, in Table I and IV.      
%
%
\begin{table}[h!]
\setlength{\arraycolsep}{50pt}
\renewcommand{\tabcolsep}{0.4pc} 
\renewcommand{\arraystretch}{1.5} 
\begin{tabular}{|c|c|c|c|c|}
\hline
 Evolution & Parameterisation    &$\chi^2/{\rm points} \quad e^+e^-$ 
&$\chi^2/{\rm points} \quad {\rm SIDIS}$   & $\chi^2/{\rm d.o.f.}$ \\
\hline
\hline
Standard & Standard & 0.71 & 0.89 & 0.84 \\
Standard & Polynomial & 0.83 & 0.94 & 0.92 \\
Transversity & Standard & 1.17 & 1.15 & 1.20 \\
Transversity & Polynomial & 1.02 & 0.93 & 1.00 \\
\hline
\end{tabular}
\caption{Values of $\chi ^2_{\rm d.o.f.}$ for different evolutions and parameterisations 
of the Collins function.
Separate values for $e^+e^-$  and SIDIS  data are also given.}
\label{tab:evos}
\end{table}
%

\begin{table}[h!]
\renewcommand{\tabcolsep}{0.4pc} 
\renewcommand{\arraystretch}{1.5} 
\begin{tabular}{|l|l|}
\hline
 $ N_{u_v}^{T} =0.58^{+0.42}_{-0.27} $ & ~$N_{d_v}^{T} =-1.00^{+2.00}_{-0.00}$ \\
 $ \alpha =0.79^{+1.41}_{-0.62} $ & ~$\beta =1.44^{+7.92}_{-1.42}$ \\
\hline \hline
 $ a_{\rm fav} =-0.02^{+0.07}_{-0.09} $ & ~$b_{\rm fav} =0.66^{+0.14}_{-0.12}$ \\
 $ a_{\rm dis} =-1.00^{+0.13}_{-0.00} $ & ~$b_{\rm dis} =0.12^{+0.38}_{-0.43}$ \\
 $ M^2_C  =0.27^{+0.17}_{-0.08} \textrm{ GeV}^2 $ &  \\
 \hline
\end{tabular}
\caption{
Best fit values of the 9 free parameters fixing the $u$ and $d$ valence quark
transversity distribution functions and the favoured and
disfavoured Collins fragmentation functions, as obtained by fitting
simultaneously SIDIS data on the Collins asymmetry and Belle and BaBar data on
$A_{0}^{UL}$ and $A_{0}^{UC}$, using the transversity kernel evolution and the polynomial parameterisation.}
\label{tab:param-evot}
\end{table}

%
%
%

\newpage

\section{BESIII azimuthal correlations \label{BESIII}}

Quite recently, the BESIII Collaboration have released their measurements
of the azimuthal Collins correlations in $e^+e^-$ annihilations into pion
pairs, completely analogous to those of BaBar and Belle, but at the
lower energy  $\sqrt{s} = Q = 3.65$
GeV~\cite{Ablikim:2015sma}.
BESIII has no clear jet event
shape to help reconstructing the thrust axis (i.e.~to separate hadrons coming
from different fragmenting quarks or antiquarks). In fact, the BESIII Collaboration
does not present $A_{12}$-type asymmetries. Instead, a cut on the opening
angle ($> 120^\circ$) is required to select back-to-back pion pairs; the azimuthal
correlations are then analysed in the hadronic frame, as explained in
Section~\ref{Form}.
We do not include these data in our fitting procedure. However, it is interesting
to check the description of these new sets of
measurements that our model can provide. Their low $Q^2$ values, as compared with Belle and BaBar
experiments, might help in assessing the importance of TMD evolution effects.


%

\begin{figure}[h!t]
\includegraphics[width=8.0truecm,angle=0]{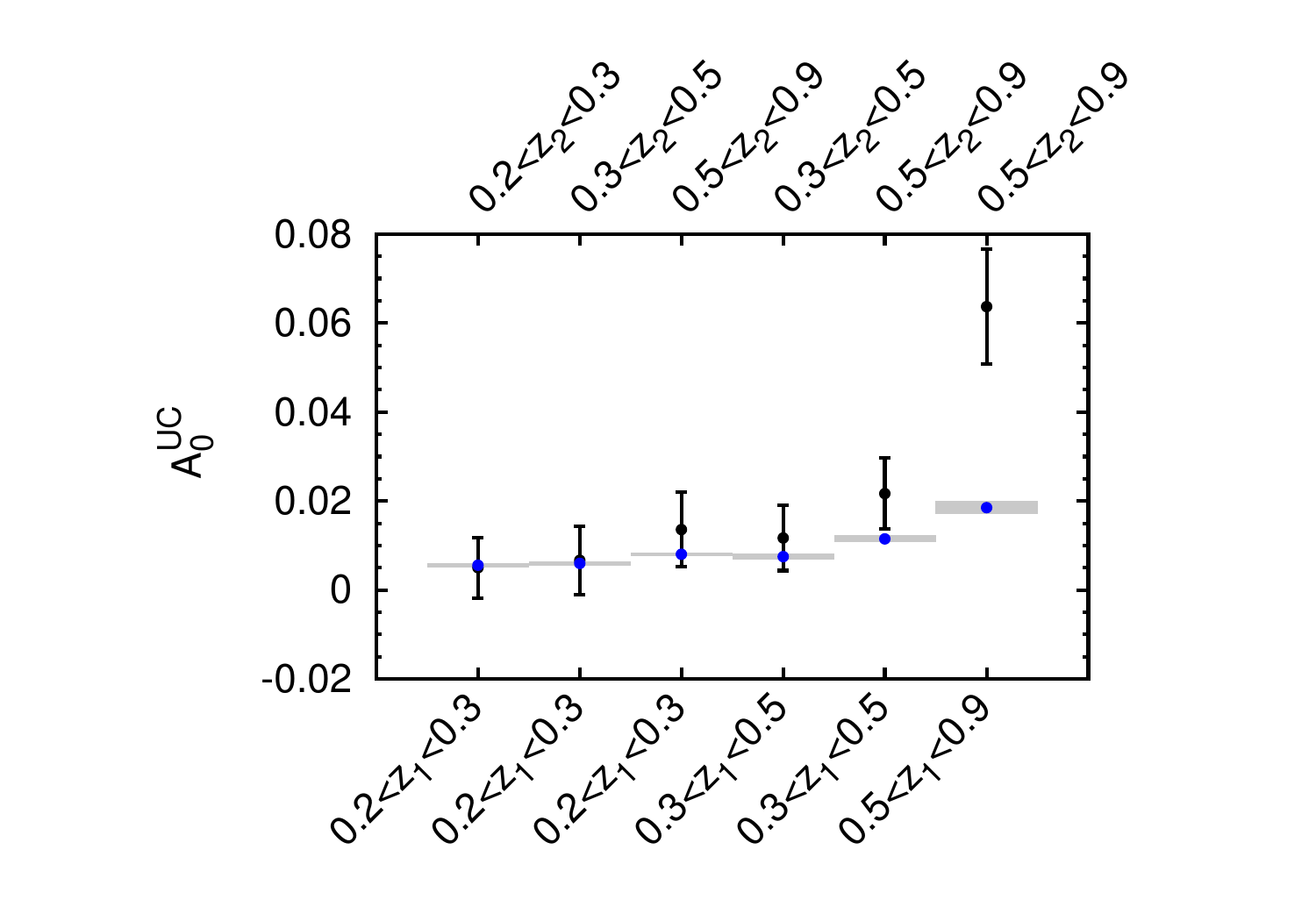}
\includegraphics[width=8.0truecm,angle=0]{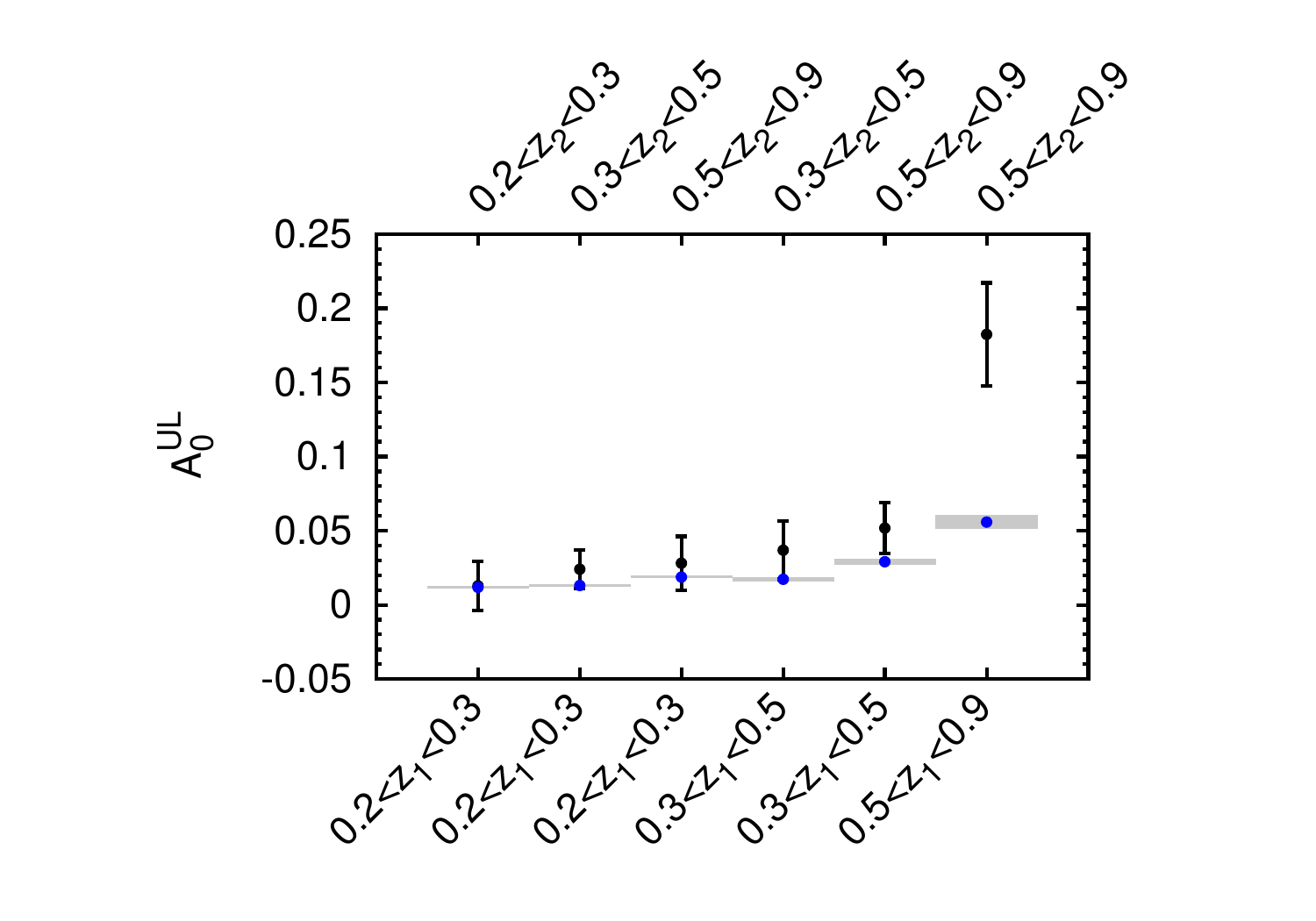}
\caption{The solid, black circles represent the $A_0^{UC}$ (left panel) and $A_0^{UL}$ (right panel)
asymmetries measured by the BESIII collaboration at $Q^2=13$ GeV$^2$, in bins
of $(z_1, z_2)$~\cite{Ablikim:2015sma}, while the solid blue circles (with their relative bands)
correspond to the predictions obtained by using our reference fit results
for the Collins functions.}
\label{fig:besIII-z}
\end{figure}

In Fig.~\ref{fig:besIII-z} the solid, black circles represent the $A_0^{UC}$
and $A_0^{UL}$ asymmetries measured by the BESIII Collaboration at $Q^2=13$
GeV$^2$, in bins of $(z_1, z_2)$, while the solid blue circles (with their
relative bands) correspond to the predictions obtained by using our reference
fit results, presented in Section~\ref{Mod}. These asymmetries are well
reproduced at small $z_1$ and $z_2$, where we expect our model to work, while
they are underestimated at very large values of either $z_1$  or $z_2$, or
both. Notice that the values of $z_1, z_2$ in the last bins are very large
for an experiment with $\sqrt{s}=3.65$ GeV: such data points might be
affected by exclusive production contributions, and other effects which
cannot be reproduced by a TMD model.

\begin{figure}[h!t]
\includegraphics[width=7.0truecm,angle=0]{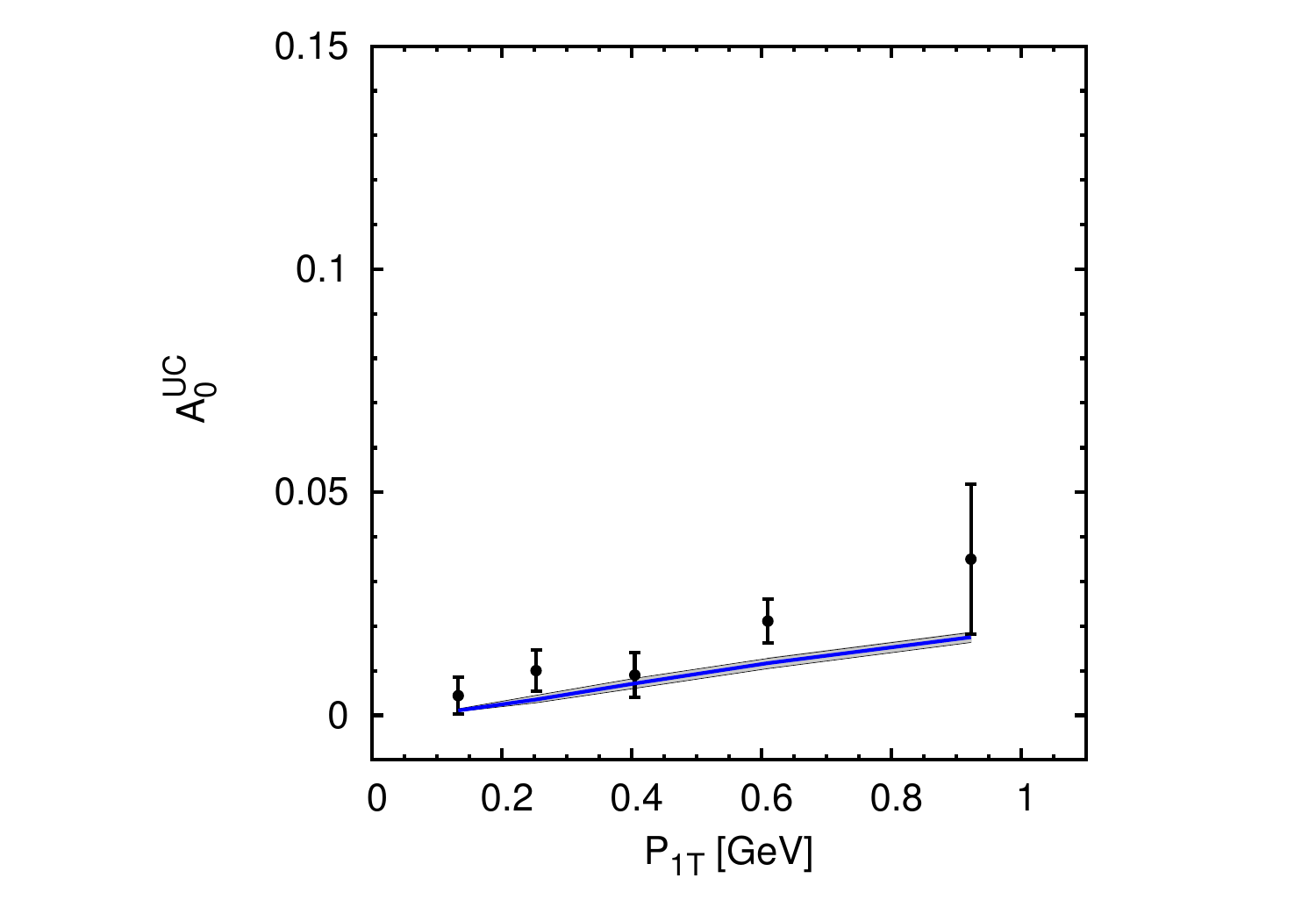}
\includegraphics[width=7.0truecm,angle=0]{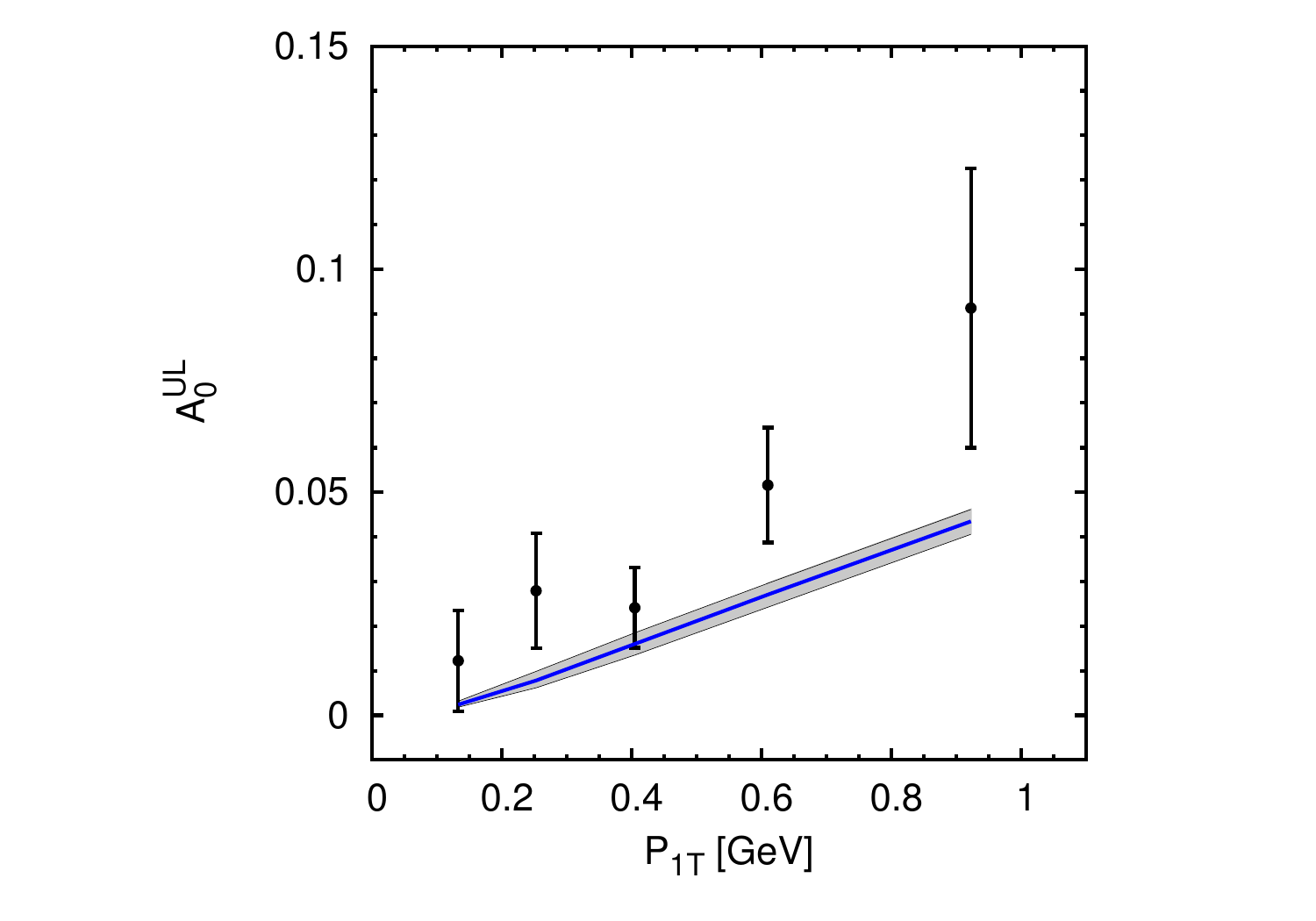}
\caption{The predictions obtained by using the Collins functions extracted from
our reference fit of SIDIS ($Q^2=2-3$ GeV$^2$) and $e^+e^-$ ($Q^2=112$ GeV$^2$)
data (solid, blue lines) are compared to the $A_0^{UC}$ (left panel) and $A_0^{UL}$ (right panel) asymmetries
measured by the BESIII collaboration~\cite{Ablikim:2015sma} at $Q^2=13$ GeV$^2$, as functions of $P_{1T}$
(black circles).
The shaded areas on the theoretical curves correspond to the
uncertainty on the parameters, as explained in the text.}
\label{fig:besIII-pt0}
\end{figure}

Fig.~\ref{fig:besIII-pt0} shows the same asymmetries, plotted as functions
of $P_{1T}$. The $A_0^{UC}$ asymmetry is described reasonably well by our
model, while $A_0^{UL}$ is slightly underestimated, especially at large
$P_{1T}$ where the effects of the experimental cuts, namely the opening
angle, become more important.

\begin{figure}[h!t]
\includegraphics[width=8.0truecm,angle=0]{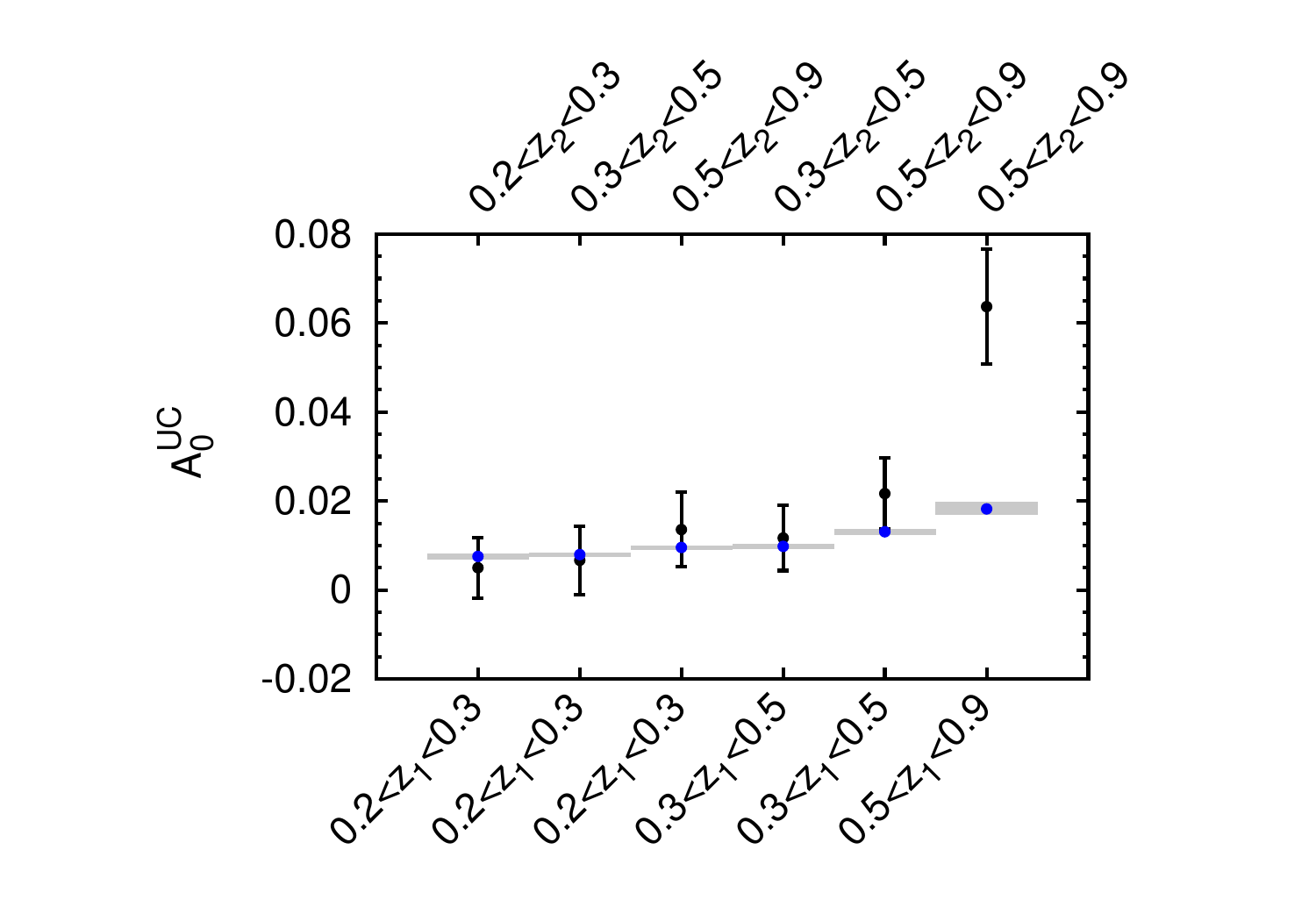}
\includegraphics[width=8.0truecm,angle=0]{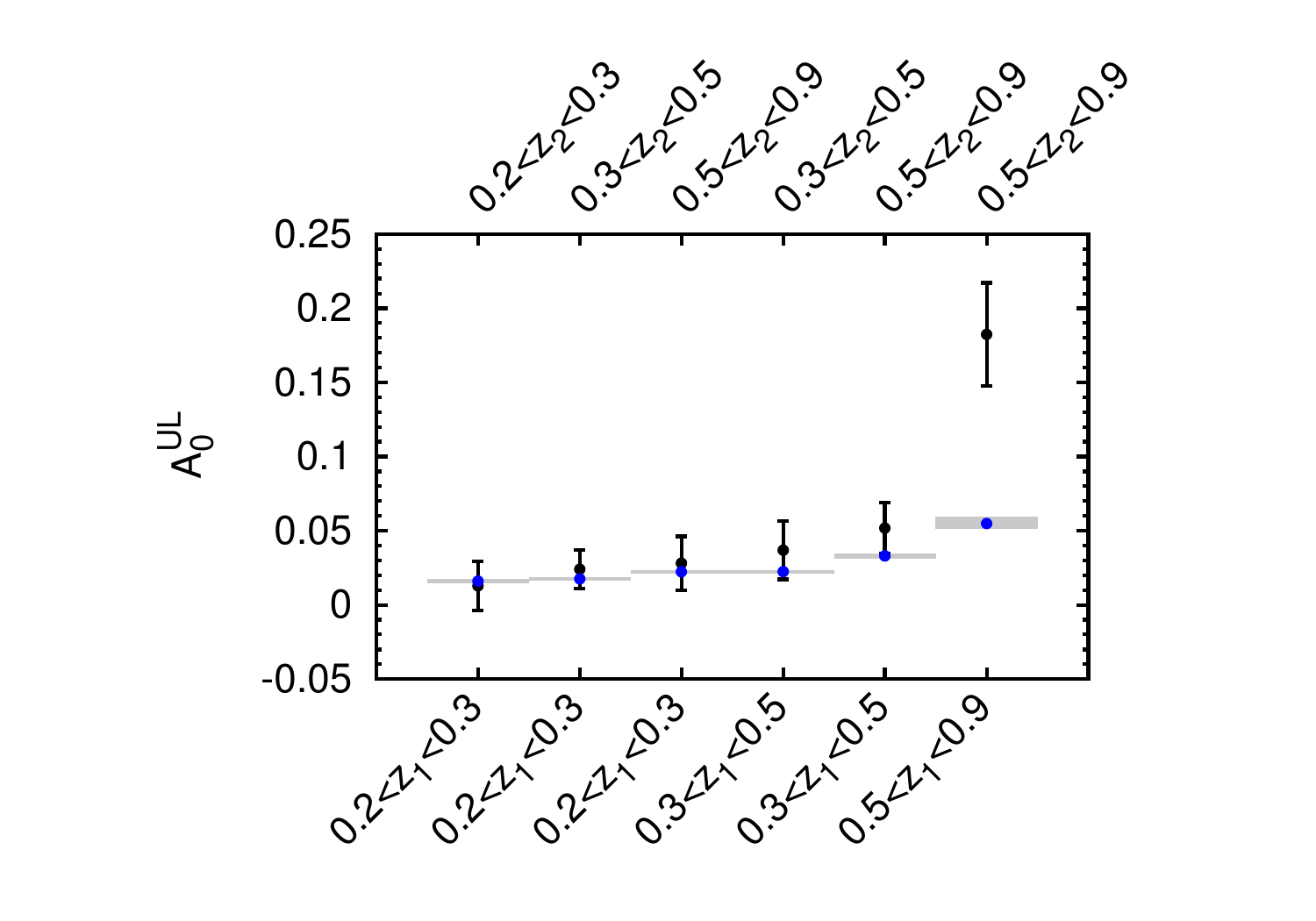}
\caption{The solid, black circles represent the $A_0^{UC}$ (left panel) and $A_0^{UL}$ (right panel)
asymmetries measured by the BESIII collaboration~\cite{Ablikim:2015sma} at $Q^2=13$ GeV$^2$, in bins
of $(z_1, z_2)$, while the solid blue circles (with their relative bands)
correspond to the predictions obtained by using the Collins functions
from our alternative fit, Table~\ref{tab:param-evot}.}
\label{fig:besIII-z-evot}
\end{figure}

\begin{figure}[h!t]
\includegraphics[width=7.0truecm,angle=0]{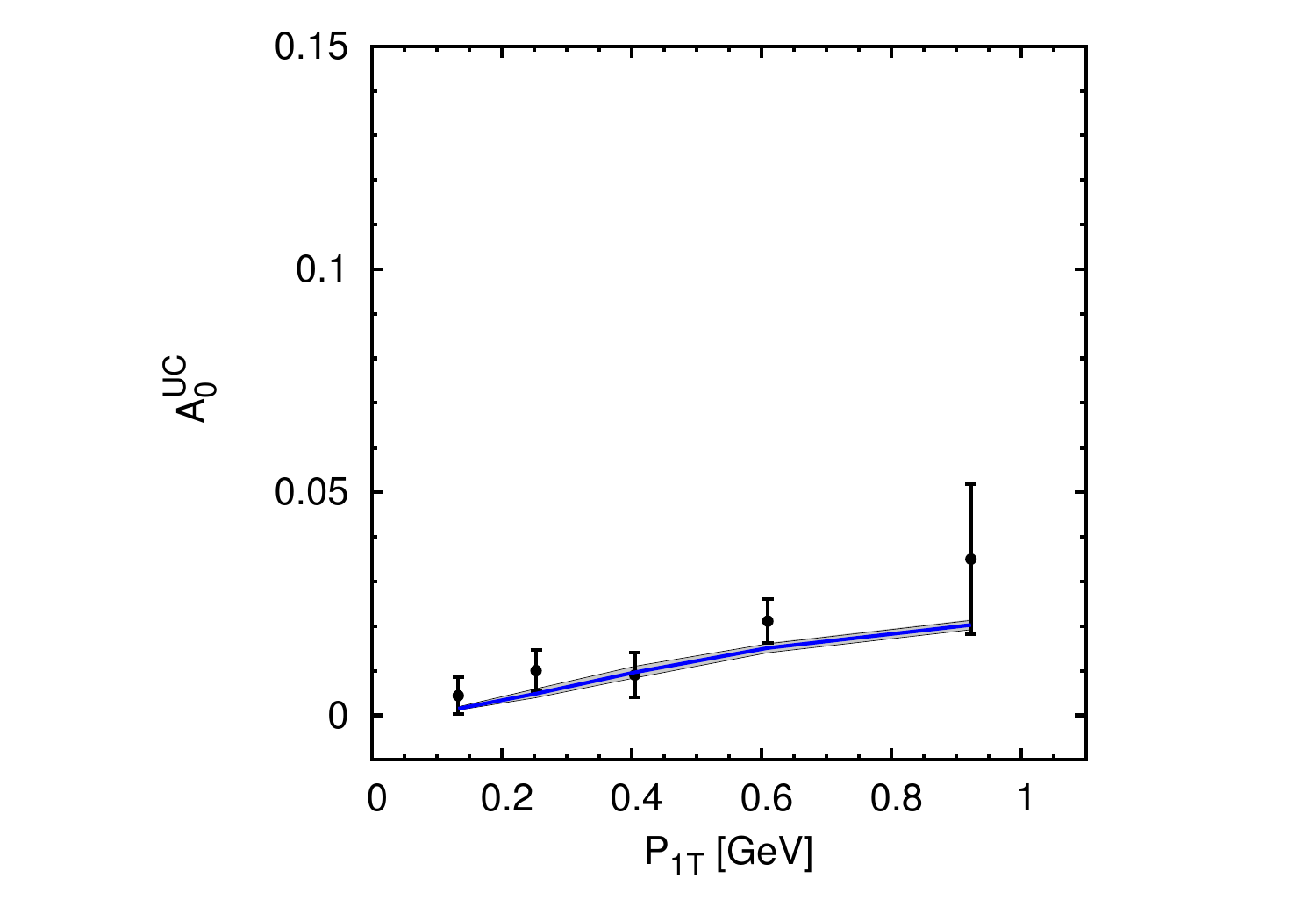}
\includegraphics[width=7.0truecm,angle=0]{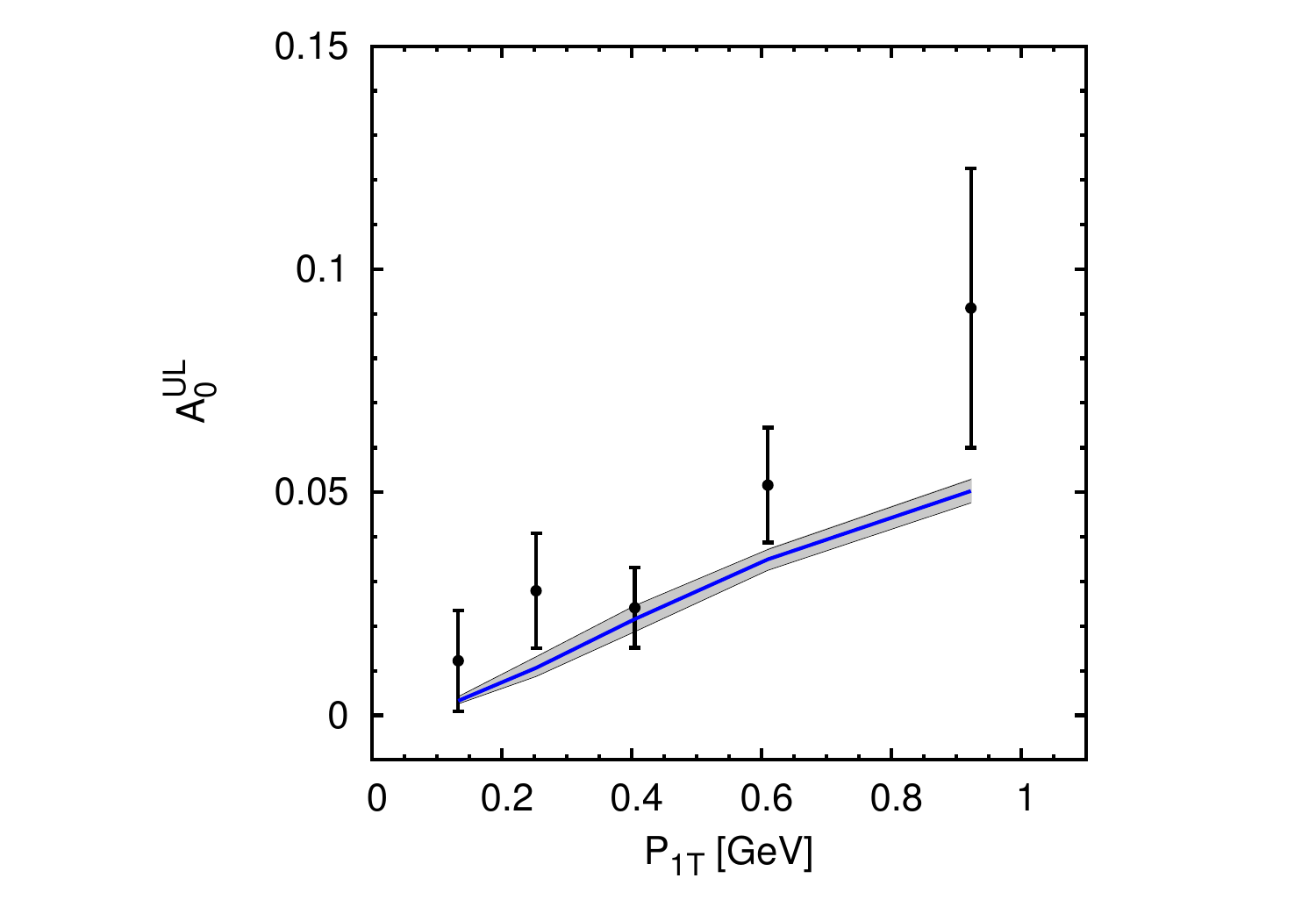}
\caption{The predictions obtained by using the Collins functions from
our alternative fit, Table~\ref{tab:param-evot},
(solid, blue lines) are compared to the $A_0^{UC}$ (left panel) and $A_0^{UL}$ (right panel) asymmetries
measured by the BESIII collaboration~\cite{Ablikim:2015sma} at $Q^2=13$ GeV$^2$, as functions of
$P_{1T}$ (black circles).
The shaded areas on the theoretical curves correspond to the
uncertainty on the parameters, as explained in the text.}
\label{fig:besIII-pt0-evot}
\end{figure}

Similar results, even with a slightly better agreement,
are obtained using the results of our alternative fit, Table~\ref{tab:param-evot},
based on a transversity evolution kernel for the Collins function combined with
a polynomial parameterisation. They are
shown in Figs.~\ref{fig:besIII-z-evot} and~\ref{fig:besIII-pt0-evot}.

At this stage, it is quite difficult to draw any clear-cut conclusion.
The predictions of our approach, which does not include TMD evolution,
seems to be quite satisfactory. On the other hand, the TMD evolution
approach of Ref.~\cite{Kang:2015msa} gives very good results.
Despite the sizeable difference in $Q^2$ among the different sets of
$e^+e^-$ data, the measured asymmetries do not show any sensitivity to evolution effects
in $Q^2$.
Further comments will be given in the conclusions.

One should also add that, at the moderate energies of BESIII experiment,
with the difficulties to isolate opposite jet hadrons, some corrections
to the TMD factorised approach might still be relevant, like the
appropriate insertion of kinematical cuts, of higher twist contributions
and of threshold effects.

%
%
%
%
%
%
%
%
%
%

\newpage

\section{Comments and conclusions}\label{Com}

We have performed a new global analysis of SIDIS and $e^+e^-$ azimuthal
asymmetries, motivated by the recent release of BaBar data, with high statistics
and precision, which offer new insights on the $p_\perp$ dependence of the Collins azimuthal 
correlations $A_0$ and $A_{12}$. 
We have extracted the Collins functions and the transversity
distributions by adopting a simple phenomenological model for these TMD--PDFs
and FFs, such that their $x$ or $z$-dependent parts evolve with $Q^2$ while the transverse
momentum dependent part is assumed to be $Q^2$ independent, i.e. by neglecting the TMD evolution.

The $u$ and $d$ quark transversity functions obtained by best fitting SIDIS
results and the new $e^+e^-$ data simultaneously are compatible with the previous
extractions~\cite{Anselmino:2007fs, Anselmino:2008jk, Anselmino:2013vqa};
while the $u$ valence transversity distribution has a clear trend, the
$d$ valence transversity still shows large uncertainties.
A similar procedure for the extraction of the transversity distributions,
which combines SIDIS and $e^+e^-$ data, involving the di-hadron fragmentation 
functions, has been adopted in
Refs.~\cite{Bacchetta:2011ip, Courtoy:2012ry, Bacchetta:2012ty}; the two
methods obtain values of the transversity distributions which are well
consistent with each other.

Instead, our newly extracted Collins functions look somewhat different from
those obtained in our previous analyses. This is mainly due to the fact that
we have exploited a different parameterisation for the disfavoured Collins
function: while in the past we used a disfavoured parameterisation with the
same shape of its favoured counterpart, but with a different normalisation
(and sign), we have now modelled the disfavoured Collins function
independently.
We have realised that one free parameter for the disfavoured Collins function 
is enough to reach a fit of excellent quality, indicating that the actual 
shape of the disfavored Collins function is still largely unconstrained by data.

About the $\pp$ dependence of the Collins function, we observe that its 
Gaussian width can now be determined with better precision.
However, our extraction is still subject to a number of initial assumptions: 
a Gaussian shape for the TMDs, a complete separation
between transverse and longitudinal degrees of freedom, a Gaussian width of 
the unpolarised TMD--FFs fixed solely by SIDIS data. 
Hopefully, higher statistics and higher precision multidimensional data, 
for asymmetries and unpolarised multiplicities,
will help clarifying the picture.

We have also made an attempt to understand the $Q^2$ dependence of these 
experimental data.
We see that our model provides a very satisfactory description of the
data and, although it relies on a $Q^2$ independent $\pp$ distribution, the 
quality of our best fit is similar to that obtained by using
TMD evolution~\cite{Kang:2015msa}.
This can be an indication that there might be cancellations of the $Q^2$ dependence
of the TMDs in these azimuthal asymmetries, which are ratios or even double
ratios of cross sections.

One can study these $Q^2$ evolution effects by directly comparing the same 
azimuthal correlations measured at very different $Q^2$ values by BaBar--Belle 
and BESIII Collaborations. Our model predicts almost identical asymmetries for 
different $Q^2$. Differences among BESIII and BaBar-Belle asymmetries
could be explained by the different kinematical configurations and cuts.
Our predictions are in qualitatively good agreement with the present BESIII 
measurements,
indicating that the data themselves do not show any strong sensitivity to 
the $Q^2$ dependence in the transverse
momentum distribution.
Also in this case, the predictions obtained from a TMD evolution approach can 
describe the data well: this points again
to cancellations of the TMD evolution effects which occur in the ratios when 
computing the measured asymmetries.

We are thus led to believe that asymmetries
or any observable which is constructed by taking ratios are not ideal grounds 
for the study of TMD evolution effects.
More effort should be made towards measuring properly normalized SIDIS and $e^+e^-$, 
and Drell-Yan cross sections (both unpolarised and polarised)
where details of TMD evolution might finally be unraveled.

\acknowledgments
\noindent
M.A., M.B., J.O.G.~and S.M.~acknowledge support from the ``Progetto di Ricerca Ateneo/CSP"
(codice TO-Call3-2012-0103). \\
U. D.~is grateful to the Department of Theoretical Physics II of the Universidad Complutense
of Madrid for the kind hospitality extended to him during the completion of this work.

\bibliography{sample}

\begin{thebibliography}{47}
\expandafter\ifx\csname natexlab\endcsname\relax\def\natexlab#1{#1}\fi
\expandafter\ifx\csname bibnamefont\endcsname\relax
  \def\bibnamefont#1{#1}\fi
\expandafter\ifx\csname bibfnamefont\endcsname\relax
  \def\bibfnamefont#1{#1}\fi
\expandafter\ifx\csname citenamefont\endcsname\relax
  \def\citenamefont#1{#1}\fi
\expandafter\ifx\csname url\endcsname\relax
  \def\url#1{\texttt{#1}}\fi
\expandafter\ifx\csname urlprefix\endcsname\relax\def\urlprefix{URL }\fi
\providecommand{\bibinfo}[2]{#2}
\providecommand{\eprint}[2][]{\url{#2}}

\bibitem[{\citenamefont{Sivers}(1990)}]{Sivers:1989cc}
\bibinfo{author}{\bibfnamefont{D.~W.} \bibnamefont{Sivers}},
  \bibinfo{journal}{Phys. Rev.} \textbf{\bibinfo{volume}{D41}},
  \bibinfo{pages}{83} (\bibinfo{year}{1990}).

\bibitem[{\citenamefont{Sivers}(1991)}]{Sivers:1990fh}
\bibinfo{author}{\bibfnamefont{D.~W.} \bibnamefont{Sivers}},
  \bibinfo{journal}{Phys. Rev.} \textbf{\bibinfo{volume}{D43}},
  \bibinfo{pages}{261} (\bibinfo{year}{1991}).

\bibitem[{\citenamefont{Collins}(1993)}]{Collins:1992kk}
\bibinfo{author}{\bibfnamefont{J.~C.} \bibnamefont{Collins}},
  \bibinfo{journal}{Nucl. Phys.} \textbf{\bibinfo{volume}{B396}},
  \bibinfo{pages}{161} (\bibinfo{year}{1993}).

\bibitem[{\citenamefont{Airapetian et~al.}(2010)}]{Airapetian:2010ds}
\bibinfo{author}{\bibfnamefont{A.}~\bibnamefont{Airapetian}}
  \bibnamefont{et~al.} (\bibinfo{collaboration}{HERMES Collaboration}),
  \bibinfo{journal}{Phys. Lett.} \textbf{\bibinfo{volume}{B693}},
  \bibinfo{pages}{11} (\bibinfo{year}{2010}), \eprint{1006.4221}.

\bibitem[{\citenamefont{Adolph et~al.}(2012)}]{Adolph:2012sn}
\bibinfo{author}{\bibfnamefont{C.}~\bibnamefont{Adolph}} \bibnamefont{et~al.}
  (\bibinfo{collaboration}{COMPASS Collaboration}), \bibinfo{journal}{Phys.
  Lett.} \textbf{\bibinfo{volume}{B717}}, \bibinfo{pages}{376}
  (\bibinfo{year}{2012}), \eprint{1205.5121}.

\bibitem[{\citenamefont{Adolph et~al.}(2015)}]{Adolph:2014zba}
\bibinfo{author}{\bibfnamefont{C.}~\bibnamefont{Adolph}} \bibnamefont{et~al.}
  (\bibinfo{collaboration}{COMPASS Collaboration}), \bibinfo{journal}{Phys.
  Lett.} \textbf{\bibinfo{volume}{B744}}, \bibinfo{pages}{250}
  (\bibinfo{year}{2015}), \eprint{1408.4405}.

\bibitem[{\citenamefont{Boer et~al.}(1997)\citenamefont{Boer, Jakob, and
  Mulders}}]{Boer:1997mf}
\bibinfo{author}{\bibfnamefont{D.}~\bibnamefont{Boer}},
  \bibinfo{author}{\bibfnamefont{R.}~\bibnamefont{Jakob}}, \bibnamefont{and}
  \bibinfo{author}{\bibfnamefont{P.}~\bibnamefont{Mulders}},
  \bibinfo{journal}{Nucl. Phys.} \textbf{\bibinfo{volume}{B504}},
  \bibinfo{pages}{345} (\bibinfo{year}{1997}), \eprint{hep-ph/9702281}.

\bibitem[{\citenamefont{Abe et~al.}(2006)}]{Abe:2005zx}
\bibinfo{author}{\bibfnamefont{K.}~\bibnamefont{Abe}} \bibnamefont{et~al.}
  (\bibinfo{collaboration}{Belle Collaboration}), \bibinfo{journal}{Phys. Rev.
  Lett.} \textbf{\bibinfo{volume}{96}}, \bibinfo{pages}{232002}
  (\bibinfo{year}{2006}), \eprint{hep-ex/0507063}.

\bibitem[{\citenamefont{Anselmino et~al.}(2007)\citenamefont{Anselmino,
  Boglione, D'Alesio, Kotzinian, Murgia et~al.}}]{Anselmino:2007fs}
\bibinfo{author}{\bibfnamefont{M.}~\bibnamefont{Anselmino}},
  \bibinfo{author}{\bibfnamefont{M.}~\bibnamefont{Boglione}},
  \bibinfo{author}{\bibfnamefont{U.}~\bibnamefont{D'Alesio}},
  \bibinfo{author}{\bibfnamefont{A.}~\bibnamefont{Kotzinian}},
  \bibinfo{author}{\bibfnamefont{F.}~\bibnamefont{Murgia}},
  \bibnamefont{et~al.}, \bibinfo{journal}{Phys. Rev.}
  \textbf{\bibinfo{volume}{D75}}, \bibinfo{pages}{054032}
  (\bibinfo{year}{2007}), \eprint{hep-ph/0701006}.

\bibitem[{\citenamefont{Anselmino
  et~al.}(2009{\natexlab{a}})\citenamefont{Anselmino, Boglione, D'Alesio,
  Kotzinian, Murgia et~al.}}]{Anselmino:2008jk}
\bibinfo{author}{\bibfnamefont{M.}~\bibnamefont{Anselmino}},
  \bibinfo{author}{\bibfnamefont{M.}~\bibnamefont{Boglione}},
  \bibinfo{author}{\bibfnamefont{U.}~\bibnamefont{D'Alesio}},
  \bibinfo{author}{\bibfnamefont{A.}~\bibnamefont{Kotzinian}},
  \bibinfo{author}{\bibfnamefont{F.}~\bibnamefont{Murgia}},
  \bibnamefont{et~al.}, \bibinfo{journal}{Nucl. Phys. Proc. Suppl.}
  \textbf{\bibinfo{volume}{191}}, \bibinfo{pages}{98}
  (\bibinfo{year}{2009}{\natexlab{a}}), \eprint{0812.4366}.

\bibitem[{\citenamefont{Anselmino
  et~al.}(2013{\natexlab{a}})\citenamefont{Anselmino, Boglione, D'Alesio,
  Melis, Murgia et~al.}}]{Anselmino:2013vqa}
\bibinfo{author}{\bibfnamefont{M.}~\bibnamefont{Anselmino}},
  \bibinfo{author}{\bibfnamefont{M.}~\bibnamefont{Boglione}},
  \bibinfo{author}{\bibfnamefont{U.}~\bibnamefont{D'Alesio}},
  \bibinfo{author}{\bibfnamefont{S.}~\bibnamefont{Melis}},
  \bibinfo{author}{\bibfnamefont{F.}~\bibnamefont{Murgia}},
  \bibnamefont{et~al.}, \bibinfo{journal}{Phys. Rev.}
  \textbf{\bibinfo{volume}{D87}}, \bibinfo{pages}{094019}
  (\bibinfo{year}{2013}{\natexlab{a}}), \eprint{1303.3822}.

\bibitem[{\citenamefont{Lees et~al.}(2014)}]{TheBABAR:2013yha}
\bibinfo{author}{\bibfnamefont{J.}~\bibnamefont{Lees}} \bibnamefont{et~al.}
  (\bibinfo{collaboration}{BaBar Collaboration}), \bibinfo{journal}{Phys. Rev.}
  \textbf{\bibinfo{volume}{D90}}, \bibinfo{pages}{052003}
  (\bibinfo{year}{2014}), \eprint{1309.5278}.

\bibitem[{\citenamefont{Schweitzer et~al.}(2010)\citenamefont{Schweitzer,
  Teckentrup, and Metz}}]{Schweitzer:2010tt}
\bibinfo{author}{\bibfnamefont{P.}~\bibnamefont{Schweitzer}},
  \bibinfo{author}{\bibfnamefont{T.}~\bibnamefont{Teckentrup}},
  \bibnamefont{and} \bibinfo{author}{\bibfnamefont{A.}~\bibnamefont{Metz}},
  \bibinfo{journal}{Phys. Rev.} \textbf{\bibinfo{volume}{D81}},
  \bibinfo{pages}{094019} (\bibinfo{year}{2010}), \eprint{1003.2190}.

\bibitem[{\citenamefont{Anselmino et~al.}(2014)\citenamefont{Anselmino,
  Boglione, Gonzalez~Hernandez, Melis, and Prokudin}}]{Anselmino:2013lza}
\bibinfo{author}{\bibfnamefont{M.}~\bibnamefont{Anselmino}},
  \bibinfo{author}{\bibfnamefont{M.}~\bibnamefont{Boglione}},
  \bibinfo{author}{\bibfnamefont{J.}~\bibnamefont{Gonzalez~Hernandez}},
  \bibinfo{author}{\bibfnamefont{S.}~\bibnamefont{Melis}}, \bibnamefont{and}
  \bibinfo{author}{\bibfnamefont{A.}~\bibnamefont{Prokudin}},
  \bibinfo{journal}{JHEP} \textbf{\bibinfo{volume}{1404}}, \bibinfo{pages}{005}
  (\bibinfo{year}{2014}), \eprint{1312.6261}.

\bibitem[{\citenamefont{Signori et~al.}(2013)\citenamefont{Signori, Bacchetta,
  Radici, and Schnell}}]{Signori:2013mda}
\bibinfo{author}{\bibfnamefont{A.}~\bibnamefont{Signori}},
  \bibinfo{author}{\bibfnamefont{A.}~\bibnamefont{Bacchetta}},
  \bibinfo{author}{\bibfnamefont{M.}~\bibnamefont{Radici}}, \bibnamefont{and}
  \bibinfo{author}{\bibfnamefont{G.}~\bibnamefont{Schnell}},
  \bibinfo{journal}{JHEP} \textbf{\bibinfo{volume}{1311}}, \bibinfo{pages}{194}
  (\bibinfo{year}{2013}), \eprint{1309.3507}.

\bibitem[{\citenamefont{Anselmino
  et~al.}(2009{\natexlab{b}})\citenamefont{Anselmino, Boglione, D'Alesio,
  Kotzinian, Melis, Murgia, Prokudin, and Turk}}]{Anselmino:2008sga}
\bibinfo{author}{\bibfnamefont{M.}~\bibnamefont{Anselmino}},
  \bibinfo{author}{\bibfnamefont{M.}~\bibnamefont{Boglione}},
  \bibinfo{author}{\bibfnamefont{U.}~\bibnamefont{D'Alesio}},
  \bibinfo{author}{\bibfnamefont{A.}~\bibnamefont{Kotzinian}},
  \bibinfo{author}{\bibfnamefont{S.}~\bibnamefont{Melis}},
  \bibinfo{author}{\bibfnamefont{F.}~\bibnamefont{Murgia}},
  \bibinfo{author}{\bibfnamefont{A.}~\bibnamefont{Prokudin}}, \bibnamefont{and}
  \bibinfo{author}{\bibfnamefont{C.}~\bibnamefont{Turk}},
  \bibinfo{journal}{Eur. Phys. J.} \textbf{\bibinfo{volume}{A39}},
  \bibinfo{pages}{89} (\bibinfo{year}{2009}{\natexlab{b}}), \eprint{0805.2677}.

\bibitem[{\citenamefont{Anselmino
  et~al.}(2012{\natexlab{a}})\citenamefont{Anselmino, Boglione, and
  Melis}}]{Anselmino:2012aa}
\bibinfo{author}{\bibfnamefont{M.}~\bibnamefont{Anselmino}},
  \bibinfo{author}{\bibfnamefont{M.}~\bibnamefont{Boglione}}, \bibnamefont{and}
  \bibinfo{author}{\bibfnamefont{S.}~\bibnamefont{Melis}},
  \bibinfo{journal}{Phys. Rev.} \textbf{\bibinfo{volume}{D86}},
  \bibinfo{pages}{014028} (\bibinfo{year}{2012}{\natexlab{a}}),
  \eprint{1204.1239}.

\bibitem[{\citenamefont{Anselmino
  et~al.}(2012{\natexlab{b}})\citenamefont{Anselmino, Boglione, D'Alesio,
  Leader, Melis et~al.}}]{Anselmino:2012rq}
\bibinfo{author}{\bibfnamefont{M.}~\bibnamefont{Anselmino}},
  \bibinfo{author}{\bibfnamefont{M.}~\bibnamefont{Boglione}},
  \bibinfo{author}{\bibfnamefont{U.}~\bibnamefont{D'Alesio}},
  \bibinfo{author}{\bibfnamefont{E.}~\bibnamefont{Leader}},
  \bibinfo{author}{\bibfnamefont{S.}~\bibnamefont{Melis}},
  \bibnamefont{et~al.}, \bibinfo{journal}{Phys. Rev.}
  \textbf{\bibinfo{volume}{D86}}, \bibinfo{pages}{074032}
  (\bibinfo{year}{2012}{\natexlab{b}}), \eprint{1207.6529}.

\bibitem[{\citenamefont{Anselmino
  et~al.}(2013{\natexlab{b}})\citenamefont{Anselmino, Boglione, D'Alesio,
  Melis, Murgia et~al.}}]{Anselmino:2013rya}
\bibinfo{author}{\bibfnamefont{M.}~\bibnamefont{Anselmino}},
  \bibinfo{author}{\bibfnamefont{M.}~\bibnamefont{Boglione}},
  \bibinfo{author}{\bibfnamefont{U.}~\bibnamefont{D'Alesio}},
  \bibinfo{author}{\bibfnamefont{S.}~\bibnamefont{Melis}},
  \bibinfo{author}{\bibfnamefont{F.}~\bibnamefont{Murgia}},
  \bibnamefont{et~al.}, \bibinfo{journal}{Phys. Rev.}
  \textbf{\bibinfo{volume}{D88}}, \bibinfo{pages}{054023}
  (\bibinfo{year}{2013}{\natexlab{b}}), \eprint{1304.7691}.

\bibitem[{\citenamefont{Collins}(2011)}]{Collins:2011qcdbook}
\bibinfo{author}{\bibfnamefont{J.~C.} \bibnamefont{Collins}},
  \emph{\bibinfo{title}{Foundations of Perturbative QCD}}
  (\bibinfo{publisher}{Cambridge University Press},
  \bibinfo{address}{Cambridge}, \bibinfo{year}{2011}).

\bibitem[{\citenamefont{Collins and Soper}(1981)}]{Collins:1981uk}
\bibinfo{author}{\bibfnamefont{J.~C.} \bibnamefont{Collins}} \bibnamefont{and}
  \bibinfo{author}{\bibfnamefont{D.~E.} \bibnamefont{Soper}},
  \bibinfo{journal}{Nucl. Phys.} \textbf{\bibinfo{volume}{B193}},
  \bibinfo{pages}{381} (\bibinfo{year}{1981}), \bibinfo{note}{[Erratum: Nucl.
  Phys.B213,545(1983)]}.

\bibitem[{\citenamefont{Collins and Soper}(1982)}]{Collins:1981uw}
\bibinfo{author}{\bibfnamefont{J.~C.} \bibnamefont{Collins}} \bibnamefont{and}
  \bibinfo{author}{\bibfnamefont{D.~E.} \bibnamefont{Soper}},
  \bibinfo{journal}{Nucl. Phys.} \textbf{\bibinfo{volume}{B194}},
  \bibinfo{pages}{445} (\bibinfo{year}{1982}).

\bibitem[{\citenamefont{Aybat and Rogers}(2011)}]{Aybat:2011zv}
\bibinfo{author}{\bibfnamefont{S.~M.} \bibnamefont{Aybat}} \bibnamefont{and}
  \bibinfo{author}{\bibfnamefont{T.~C.} \bibnamefont{Rogers}},
  \bibinfo{journal}{Phys. Rev.} \textbf{\bibinfo{volume}{D83}},
  \bibinfo{pages}{114042} (\bibinfo{year}{2011}), \eprint{1101.5057}.

\bibitem[{\citenamefont{Echevarria et~al.}(2013)\citenamefont{Echevarria,
  Idilbi, Schafer, and Scimemi}}]{Echevarria:2012pw}
\bibinfo{author}{\bibfnamefont{M.~G.} \bibnamefont{Echevarria}},
  \bibinfo{author}{\bibfnamefont{A.}~\bibnamefont{Idilbi}},
  \bibinfo{author}{\bibfnamefont{A.}~\bibnamefont{Schafer}}, \bibnamefont{and}
  \bibinfo{author}{\bibfnamefont{I.}~\bibnamefont{Scimemi}},
  \bibinfo{journal}{Eur. Phys. J.} \textbf{\bibinfo{volume}{C73}},
  \bibinfo{pages}{2636} (\bibinfo{year}{2013}), \eprint{1208.1281}.

\bibitem[{\citenamefont{Echevarria et~al.}(2014)\citenamefont{Echevarria,
  Idilbi, and Scimemi}}]{Echevarria:2014rua}
\bibinfo{author}{\bibfnamefont{M.~G.} \bibnamefont{Echevarria}},
  \bibinfo{author}{\bibfnamefont{A.}~\bibnamefont{Idilbi}}, \bibnamefont{and}
  \bibinfo{author}{\bibfnamefont{I.}~\bibnamefont{Scimemi}},
  \bibinfo{journal}{Phys. Rev.} \textbf{\bibinfo{volume}{D90}},
  \bibinfo{pages}{014003} (\bibinfo{year}{2014}), \eprint{1402.0869}.

\bibitem[{\citenamefont{Kang et~al.}(2015)\citenamefont{Kang, Prokudin, Sun,
  and Yuan}}]{Kang:2015msa}
\bibinfo{author}{\bibfnamefont{Z.-B.} \bibnamefont{Kang}},
  \bibinfo{author}{\bibfnamefont{A.}~\bibnamefont{Prokudin}},
  \bibinfo{author}{\bibfnamefont{P.}~\bibnamefont{Sun}}, \bibnamefont{and}
  \bibinfo{author}{\bibfnamefont{F.}~\bibnamefont{Yuan}}
  (\bibinfo{year}{2015}), \eprint{1505.05589}.

\bibitem[{\citenamefont{Aidala et~al.}(2014)\citenamefont{Aidala, Field,
  Gamberg, and Rogers}}]{Aidala:2014hva}
\bibinfo{author}{\bibfnamefont{C.~A.} \bibnamefont{Aidala}},
  \bibinfo{author}{\bibfnamefont{B.}~\bibnamefont{Field}},
  \bibinfo{author}{\bibfnamefont{L.~P.} \bibnamefont{Gamberg}},
  \bibnamefont{and} \bibinfo{author}{\bibfnamefont{T.~C.}
  \bibnamefont{Rogers}}, \bibinfo{journal}{Phys. Rev.}
  \textbf{\bibinfo{volume}{D89}}, \bibinfo{pages}{094002}
  (\bibinfo{year}{2014}), \eprint{1401.2654}.

\bibitem[{\citenamefont{Collins and Rogers}(2015)}]{Collins:2014jpa}
\bibinfo{author}{\bibfnamefont{J.}~\bibnamefont{Collins}} \bibnamefont{and}
  \bibinfo{author}{\bibfnamefont{T.}~\bibnamefont{Rogers}},
  \bibinfo{journal}{Phys. Rev.} \textbf{\bibinfo{volume}{D91}},
  \bibinfo{pages}{074020} (\bibinfo{year}{2015}), \eprint{1412.3820}.

\bibitem[{\citenamefont{Ablikim et~al.}(2015)}]{Ablikim:2015sma}
\bibinfo{author}{\bibfnamefont{M.}~\bibnamefont{Ablikim}} \bibnamefont{et~al.}
  (\bibinfo{collaboration}{BESIII Collaboration}) (\bibinfo{year}{2015}),
  \eprint{1507.06824}.

\bibitem[{\citenamefont{Aubert et~al.}(2015)}]{Aubert:2015hha}
\bibinfo{author}{\bibfnamefont{B.}~\bibnamefont{Aubert}} \bibnamefont{et~al.}
  (\bibinfo{collaboration}{BaBar Collaboration}) (\bibinfo{year}{2015}),
  \eprint{1506.05864}.

\bibitem[{\citenamefont{Seidl et~al.}(2008)}]{Seidl:2008xc}
\bibinfo{author}{\bibfnamefont{R.}~\bibnamefont{Seidl}} \bibnamefont{et~al.}
  (\bibinfo{collaboration}{Belle Collaboration}), \bibinfo{journal}{Phys. Rev.}
  \textbf{\bibinfo{volume}{D78}}, \bibinfo{pages}{032011}
  (\bibinfo{year}{2008}), \eprint{0805.2975}.

\bibitem[{\citenamefont{Seidl et~al.}(2012)}]{Seidl:2012er}
\bibinfo{author}{\bibfnamefont{R.}~\bibnamefont{Seidl}} \bibnamefont{et~al.}
  (\bibinfo{collaboration}{Belle Collaboration}), \bibinfo{journal}{Phys. Rev.}
  \textbf{\bibinfo{volume}{D86}}, \bibinfo{pages}{039905(E)}
  (\bibinfo{year}{2012}), \eprint{0805.2975}.

\bibitem[{\citenamefont{Martin}(2014)}]{Martin:2013eja}
\bibinfo{author}{\bibfnamefont{A.}~\bibnamefont{Martin}}
  (\bibinfo{collaboration}{COMPASS Collaboration}), \bibinfo{journal}{Phys.
  Part. Nucl.} \textbf{\bibinfo{volume}{45}}, \bibinfo{pages}{141}
  (\bibinfo{year}{2014}), \eprint{1303.2076}.

\bibitem[{\citenamefont{Anselmino et~al.}(2011)\citenamefont{Anselmino,
  Boglione, D'Alesio, Melis, Murgia et~al.}}]{Anselmino:2011ch}
\bibinfo{author}{\bibfnamefont{M.}~\bibnamefont{Anselmino}},
  \bibinfo{author}{\bibfnamefont{M.}~\bibnamefont{Boglione}},
  \bibinfo{author}{\bibfnamefont{U.}~\bibnamefont{D'Alesio}},
  \bibinfo{author}{\bibfnamefont{S.}~\bibnamefont{Melis}},
  \bibinfo{author}{\bibfnamefont{F.}~\bibnamefont{Murgia}},
  \bibnamefont{et~al.}, \bibinfo{journal}{Phys. Rev.}
  \textbf{\bibinfo{volume}{D83}}, \bibinfo{pages}{114019}
  (\bibinfo{year}{2011}), \eprint{1101.1011}.

\bibitem[{\citenamefont{Bacchetta et~al.}(2007)\citenamefont{Bacchetta, Diehl,
  Goeke, Metz, Mulders et~al.}}]{Bacchetta:2006tn}
\bibinfo{author}{\bibfnamefont{A.}~\bibnamefont{Bacchetta}},
  \bibinfo{author}{\bibfnamefont{M.}~\bibnamefont{Diehl}},
  \bibinfo{author}{\bibfnamefont{K.}~\bibnamefont{Goeke}},
  \bibinfo{author}{\bibfnamefont{A.}~\bibnamefont{Metz}},
  \bibinfo{author}{\bibfnamefont{P.~J.} \bibnamefont{Mulders}},
  \bibnamefont{et~al.}, \bibinfo{journal}{JHEP}
  \textbf{\bibinfo{volume}{0702}}, \bibinfo{pages}{093} (\bibinfo{year}{2007}),
  \eprint{hep-ph/0611265}.

\bibitem[{\citenamefont{Bacchetta et~al.}(2004)\citenamefont{Bacchetta,
  D'Alesio, Diehl, and Miller}}]{Bacchetta:2004jz}
\bibinfo{author}{\bibfnamefont{A.}~\bibnamefont{Bacchetta}},
  \bibinfo{author}{\bibfnamefont{U.}~\bibnamefont{D'Alesio}},
  \bibinfo{author}{\bibfnamefont{M.}~\bibnamefont{Diehl}}, \bibnamefont{and}
  \bibinfo{author}{\bibfnamefont{C.~A.} \bibnamefont{Miller}},
  \bibinfo{journal}{Phys. Rev.} \textbf{\bibinfo{volume}{D70}},
  \bibinfo{pages}{117504} (\bibinfo{year}{2004}), \eprint{hep-ph/0410050}.

\bibitem[{\citenamefont{Gluck et~al.}(1998)\citenamefont{Gluck, Reya, and
  Vogt}}]{Gluck:1998xa}
\bibinfo{author}{\bibfnamefont{M.}~\bibnamefont{Gluck}},
  \bibinfo{author}{\bibfnamefont{E.}~\bibnamefont{Reya}}, \bibnamefont{and}
  \bibinfo{author}{\bibfnamefont{A.}~\bibnamefont{Vogt}},
  \bibinfo{journal}{Eur. Phys. J.} \textbf{\bibinfo{volume}{C5}},
  \bibinfo{pages}{461} (\bibinfo{year}{1998}), \eprint{hep-ph/9806404}.

\bibitem[{\citenamefont{de~Florian et~al.}(2007)\citenamefont{de~Florian,
  Sassot, and Stratmann}}]{deFlorian:2007hc}
\bibinfo{author}{\bibfnamefont{D.}~\bibnamefont{de~Florian}},
  \bibinfo{author}{\bibfnamefont{R.}~\bibnamefont{Sassot}}, \bibnamefont{and}
  \bibinfo{author}{\bibfnamefont{M.}~\bibnamefont{Stratmann}},
  \bibinfo{journal}{Phys. Rev.} \textbf{\bibinfo{volume}{D76}},
  \bibinfo{pages}{074033} (\bibinfo{year}{2007}), \eprint{0707.1506}.

\bibitem[{\citenamefont{Pumplin et~al.}(2002)\citenamefont{Pumplin, Stump,
  Huston, Lai, Nadolsky, and Tung}}]{Pumplin:2002vw}
\bibinfo{author}{\bibfnamefont{J.}~\bibnamefont{Pumplin}},
  \bibinfo{author}{\bibfnamefont{D.~R.} \bibnamefont{Stump}},
  \bibinfo{author}{\bibfnamefont{J.}~\bibnamefont{Huston}},
  \bibinfo{author}{\bibfnamefont{H.~L.} \bibnamefont{Lai}},
  \bibinfo{author}{\bibfnamefont{P.~M.} \bibnamefont{Nadolsky}},
  \bibnamefont{and} \bibinfo{author}{\bibfnamefont{W.~K.} \bibnamefont{Tung}},
  \bibinfo{journal}{JHEP} \textbf{\bibinfo{volume}{07}}, \bibinfo{pages}{012}
  (\bibinfo{year}{2002}), \eprint{hep-ph/0201195}.

\bibitem[{\citenamefont{Salam and Rojo}(2009)}]{Salam:2008qg}
\bibinfo{author}{\bibfnamefont{G.~P.} \bibnamefont{Salam}} \bibnamefont{and}
  \bibinfo{author}{\bibfnamefont{J.}~\bibnamefont{Rojo}},
  \bibinfo{journal}{Comput. Phys. Commun.} \textbf{\bibinfo{volume}{180}},
  \bibinfo{pages}{120} (\bibinfo{year}{2009}), \eprint{0804.3755}.

\bibitem[{\citenamefont{Gluck et~al.}(2001)\citenamefont{Gluck, Reya,
  Stratmann, and Vogelsang}}]{Gluck:2000dy}
\bibinfo{author}{\bibfnamefont{M.}~\bibnamefont{Gluck}},
  \bibinfo{author}{\bibfnamefont{E.}~\bibnamefont{Reya}},
  \bibinfo{author}{\bibfnamefont{M.}~\bibnamefont{Stratmann}},
  \bibnamefont{and}
  \bibinfo{author}{\bibfnamefont{W.}~\bibnamefont{Vogelsang}},
  \bibinfo{journal}{Phys. Rev.} \textbf{\bibinfo{volume}{D63}},
  \bibinfo{pages}{094005} (\bibinfo{year}{2001}), \eprint{hep-ph/0011215}.

\bibitem[{\citenamefont{Yuan and Zhou}(2009)}]{Yuan:2009dw}
\bibinfo{author}{\bibfnamefont{F.}~\bibnamefont{Yuan}} \bibnamefont{and}
  \bibinfo{author}{\bibfnamefont{J.}~\bibnamefont{Zhou}},
  \bibinfo{journal}{Phys. Rev. Lett.} \textbf{\bibinfo{volume}{103}},
  \bibinfo{pages}{052001} (\bibinfo{year}{2009}), \eprint{0903.4680}.

\bibitem[{\citenamefont{Kang}(2011)}]{Kang:2010xv}
\bibinfo{author}{\bibfnamefont{Z.-B.} \bibnamefont{Kang}},
  \bibinfo{journal}{Phys. Rev.} \textbf{\bibinfo{volume}{D83}},
  \bibinfo{pages}{036006} (\bibinfo{year}{2011}), \eprint{1012.3419}.

\bibitem[{\citenamefont{Farouki}(2012)}]{Farouki:2012}
\bibinfo{author}{\bibfnamefont{R.~T.} \bibnamefont{Farouki}},
  \bibinfo{journal}{Computer Aided Geometric Design}
  \textbf{\bibinfo{volume}{29}}, \bibinfo{pages}{379} (\bibinfo{year}{2012}).

\bibitem[{\citenamefont{Bacchetta et~al.}(2011)\citenamefont{Bacchetta,
  Courtoy, and Radici}}]{Bacchetta:2011ip}
\bibinfo{author}{\bibfnamefont{A.}~\bibnamefont{Bacchetta}},
  \bibinfo{author}{\bibfnamefont{A.}~\bibnamefont{Courtoy}}, \bibnamefont{and}
  \bibinfo{author}{\bibfnamefont{M.}~\bibnamefont{Radici}},
  \bibinfo{journal}{Phys. Rev. Lett.} \textbf{\bibinfo{volume}{107}},
  \bibinfo{pages}{012001} (\bibinfo{year}{2011}), \eprint{1104.3855}.

\bibitem[{\citenamefont{Courtoy et~al.}(2012)\citenamefont{Courtoy, Bacchetta,
  Radici, and Bianconi}}]{Courtoy:2012ry}
\bibinfo{author}{\bibfnamefont{A.}~\bibnamefont{Courtoy}},
  \bibinfo{author}{\bibfnamefont{A.}~\bibnamefont{Bacchetta}},
  \bibinfo{author}{\bibfnamefont{M.}~\bibnamefont{Radici}}, \bibnamefont{and}
  \bibinfo{author}{\bibfnamefont{A.}~\bibnamefont{Bianconi}},
  \bibinfo{journal}{Phys. Rev.} \textbf{\bibinfo{volume}{D85}},
  \bibinfo{pages}{114023} (\bibinfo{year}{2012}), \eprint{1202.0323}.

\bibitem[{\citenamefont{Bacchetta et~al.}(2013)\citenamefont{Bacchetta,
  Courtoy, and Radici}}]{Bacchetta:2012ty}
\bibinfo{author}{\bibfnamefont{A.}~\bibnamefont{Bacchetta}},
  \bibinfo{author}{\bibfnamefont{A.}~\bibnamefont{Courtoy}}, \bibnamefont{and}
  \bibinfo{author}{\bibfnamefont{M.}~\bibnamefont{Radici}},
  \bibinfo{journal}{JHEP} \textbf{\bibinfo{volume}{1303}}, \bibinfo{pages}{119}
  (\bibinfo{year}{2013}), \eprint{1212.3568}.

\end{thebibliography}


\end{document}